\documentclass[a4paper,11pt]{article}
\pdfoutput=1

\usepackage{jcappub}

\usepackage[T1]{fontenc}
\usepackage{multirow}
\usepackage{amssymb}
\usepackage{mathtools}
\usepackage{color}
\usepackage{hyperref}
\usepackage{bm} 
\usepackage{graphicx}
\usepackage{xcolor}
\usepackage{lscape}
\usepackage{booktabs}
\usepackage{tabularx}
\usepackage{soul}
\newcommand{\be}{\begin{equation}}
\newcommand{\ee}{\end{equation}}
\newcommand{\nn}{\nonumber}
\newcommand{\bea}{\begin{eqnarray}}
\newcommand{\eea}{\end{eqnarray}}

\renewcommand{\d}{{\rm d}}
\newcommand{\dd}{\partial}

\newcommand{\bn}{{\bm n}}

\newcommand{\al}{\alpha}

\newcommand{\de}{\delta}
\newcommand{\De}{\Delta}
\newcommand{\ep}{\epsilon}

\newcommand{\la}{\lambda}

\newcommand{\Omm}{\Omega}

\newcommand{\ylm}[1]{Y_{\ell m}(\hat #1)}
\newcommand{\ylmc}[1]{Y^*_{\ell m}(\hat #1)}

\newcommand{\alm}{a_{\ell m}}

\newcommand{\vnd}[1]{{\color{black}{#1}}}

\newcommand{\vk}{\vec{k}}
\newcommand{\don}{{\rm d}\Omega_{\bn}}
\newcommand{\dt}{{\rm d}^3}
\renewcommand{\thefootnote}{\alph{footnote}} 
\everymath{\displaystyle}

\title{\boldmath The observed number counts in luminosity distance space}

\author[,1,2,3,4]{José Fonseca\footnote{Corresponding author.}}
\author[,3]{, Stefano Zazzera\footnote{Corresponding author.}}
\author[3]{, Tessa Baker}%\note{Also at Some University.}}
\author[3,4,5]{and Chris Clarkson}

\affiliation[1]{Instituto de Astrof\'isica e Ci\^{e}ncias do Espa\c{c}o, Universidade do Porto, CAUP, Rua das Estrelas, PT4150-762 Porto, Portugal}
\affiliation[2]{Departamento de F\'isica e Astronomia, Faculdade de Ci\^{e}ncias, Universidade do Porto, Rua do Campo Alegre 687, PT4169-007 Porto, Portugal}
\affiliation[3]{Department of Physics \& Astronomy, Queen Mary University of London, Mile End Road, London E1 4NS, United Kingdom}
\affiliation[4]{Department of Physics \& Astronomy, University of the Western Cape, Cape Town 7535, South Africa}
\affiliation[5]{Department of Mathematics \& Applied Mathematics, University of Cape Town, Cape Town 7701, South Africa}
\emailAdd{jose.fonseca@astro.up.pt}
\emailAdd{s.zazzera@qmul.ac.uk}
\emailAdd{t.baker@qmul.ac.uk}
\emailAdd{chris.clarkson@qmul.ac.uk}

\abstract{Next generation surveys will provide us with an unprecedented number of detections of supernovae Type Ia and gravitational wave merger events. Cross-correlations of such objects offer novel and powerful insights into the large-scale distribution of matter in the universe. Both of these sources carry information on their luminosity distance, but remain uninformative about their redshifts; hence their clustering analyses and cross-correlations need to be carried out in luminosity distance space, as opposed to redshift space. In this paper, we calculate the full expression for the number count fluctuation in terms of a perturbation to the observed luminosity distance. We find the expression to differ significantly from the one commonly used in redshift space. Furthermore, we present a comparison of the
number count angular power spectra between luminosity distance and redshift spaces. We see a wide divergence between the two at large scales, and we note that lensing is the main contribution to such differences. On such scales and at higher redshifts the difference between the angular power spectra in luminosity distance and redshift spaces can be roughly 50$\%$.
We also investigate cross-correlating different redshift bins using different tracers, i.e. one in luminosity distance space and one in redshift, simulating the cross-correlation angular power spectrum between background gravitational waves/supernovae and foreground galaxies. Finally, we show that in a cosmic variance limited survey, the relativistic corrections to the density-only term ought to be included.
}

\begin{document}
\maketitle
\flushbottom
%\newpage
\renewcommand{\thefootnote}{\arabic{footnote}} 
\section{Introduction}
\label{sec:intro}

Forthcoming gravitational wave experiments such as the ground-based Einstein Telescope (ET) \cite{Sathyaprakash_2010, Sathyaprakash_2012, Maggiore_2020}, and Cosmic Explorer (CE) \cite{Evans_2011}, as well as the space-based Laser Interferometer Space Antenna (LISA) \cite{LISA:2017pwj}, will detect rapidly growing numbers of gravitational wave (GW) merger events. These are detected when inspiraling compact objects, such as black holes and neutron stars, merge. Such objects are, in principle, the result of astrophysical processes in galaxies, and therefore trace the large-scale distribution of matter as their host galaxies do. However, unless an electromagnetic counterpart identifies the unique host galaxy of the merger, one can only estimate the luminosity distance to the GW source; the redshift of the event is not explicitly contained in gravitational waveform in a measurable way. 
%In general, one can only estimate the luminosity distance to the merger if no spectroscopic follow-up of the host is done.

But GW mergers are not the only observed transient events for which one estimates a luminosity distance. Supernovae Type Ia (SNIa) are another example. Although one can do a spectroscopic follow-up of the host galaxy, the redshift of the host is more commonly measured photometrically, if the host galaxy is correctly identified. Thus, the most direct measurement of the distance to a SNIa is using the distance modulus, which is dependent on the luminosity distance (and systematic effects).

Hence gravitational waves and supernovae both provide a measure of the luminosity distance of the source, but they remain agnostic on its redshift. Photometric Type Ia supernovae (SNIa) have already been established as a powerful tool for the construction of the cosmological distance ladder via their property of acting as \textit{standard candles} \cite{Riess_1998, Perlmutter_1999}. Additionally, the detection of GWs from inspiraling compact binaries can also be exploited to measure an event's luminosity distance, and as GWs can be thought of as acoustic rather than visual, a more commonly used term for them is then \textit{standard sirens} \cite{Camera_2013,Shang_2010,Taylor_2012,Holz_2005, Finn_1996, Hughes_2003, Ye_2021}. 

Further, the Legacy Survey of Space and Time (LSST) of the Vera Rubin Observatory is predicted to observe around $10^6$ photometric supernovae during its 10-year observing time \cite{Libanore2022,LSST_2021,Sanchez_2022,LSST_book},
and a similar number of GW events is forecasted to be seen by following third generation, ground-based GW observatories such as 
ET \cite{Libanore2022,Libanore2021,Scelfo_2018,Scelfo_2020,Scelfo_2022,Scelfo_2022_2,Sathyaprakash_2010,Sathyaprakash_2012,Maggiore_2020,Punturo_2010} and 
CE \cite{Evans_2011, Reitze_2019,Mpetha_2022, Evans_2021}. This unprecedented number of detected objects will allow us to go beyond using them as \textit{standard sirens} and \textit{standard candles} only. In fact, we will be able to use them to study their clustering %of these tracers 
across various scales, a method currently only applicable to galaxies \cite{Bonvin2011,Challinor_Lewis,Tansella_2018,Castorina_2022,Bonvin_2008}.

Cross-correlating gravitational waves with other large-scale structure tracers has lately been established as a potentially powerful probe of the distribution of matter in the universe, as well as in understanding the evolution of astrophysical black holes through cosmic time \cite{Libanore2022,Libanore2021,Scelfo_2018,Scelfo_2020,Scelfo_2022,Scelfo_2022_2}. However, this method will be possible only with the new generations of surveys which will provide us with an unparalleled wealth of data. In this regime, one needs to have a complete description of the observed clustering of sources for which we measure luminosity distances. One could in principle assume a cosmology and translate an observed luminosity distance into an estimate of the redshift. However, doing so imposes a strong prior on any potential constraints coming from these objects, likely %even
biasing them.

Since in General Relativity (GR) both photons and GWs follow null geodesics, both will be distorted equally by lensing caused by the foreground matter distribution \cite{Bonvin_2008,Calore_2020, Namikawa, Oguri_2018,Mukherjee_2018,Mukherjee_2020,Mukherjee_20202,Congedo_2019,You_2021}, which will impact the inferred luminosity distance to the event \cite{Bertacca2018,Hirata_2010,Cutler_2009,Camera_2013,Shang_2010,Namikawa,Libanore2022}. {Lensing is not the only physical effect that alter the observed luminosity distance, as in any perturbed FLRW universe the observed luminosity distance acquires linear corrections \cite{Bertacca2018,Sasaki:1987ad,Pyne:2003bn,Bonvin:2005ps,Hui:2005nm}. One should note that the exact expression of the perturbation depends on what one considers as the reference points, as we will discuss later.} Therefore, the observed luminosity distance is an excellent probe of large-scale structure. Analysing the angular power spectra of these tracers \cite{Namikawa,Zhang:2018nea,Hui:2005nm}, together with computing cross-correlations between different tracers that live in luminosity distance space (LDS), can be a powerful tool in constraining cosmological parameters, and further studying the distance-redshift relationship \cite{Namikawa,Arabsalmani_2013, Messenger_2012, Osato_2018, Namikawa_2016, Libanore2022}.

Similarly to the more commonly explored counterpart in redshift space (RS) \cite{Hui_2008, Chen_2008,Martinez_1999,Padmanabhan_2007, Yoo_2009, Jeong_2012}, clustering analysis in luminosity distance space has to take into account distortion effects caused by the gradient of peculiar velocities --- like the well-studied Redshift Space Distortions (RSDs), radial velocity effects, Sachs-Wolfe terms and the integrated Sachs-Wolfe (ISW) effect \cite{Bacon2014,Bonvin2011,Namikawa, Challinor_Lewis, Tansella_2018, Castorina_2022,Bonvin_2008}. While Doppler and RSD terms are dominant at low redshift, the other effects could provide significant corrections at higher distances: in particular, lensing effects are integrated over distance, thus at higher redshifts they will be expected to increase in amplitude. While LSST will be able to observe SNIa events at $z<4$ \cite{LSST_2021,Sanchez_2022,LSST_book}, ET/CE are predicted to access the mergers of stellar mass binary black holes up to $z \sim 20$ \cite{Sathyaprakash_2010,Maggiore_2020, Punturo_2010,Evans_2011}. Therefore, clustering analysis at these previously inaccessible distances would require robust calculations of the distortion effects caused by lensing, allowing for the possibility of detecting lensing effects that would otherwise be buried under low-redshift contributions.

The goal of this paper is to present the full calculation of the observed number count fluctuations of sources in luminosity distance space (LDS), including all relativistic corrections. Additionally, we developed a code to compute the angular power spectrum in LDS. Our general results can be applied to any tracer in LDS, such as GWs or SNIa. We will then use our results to assess the relative importance of the various contributions to the angular power spectrum LDS tracers, including cross-correlations with galaxy surveys. By this means, we construct a theoretical formalism applicable to data coming from future surveys like LSST and ET. Forecasts on the observability of these auto- and cross-correlations, and in particular the detectability of the various contributions is outside the scope of this paper and will be treated in future work. Similarly, forecast cosmological constraints using catalogues of LDS objects are beyond the scope of the current paper.

The paper is structured as follows. In section \ref{sec:contrast}, we compute the number density contrast in terms of the luminosity distance and volume perturbations. We also clarify the expression for the total perturbation in the observed luminosity distance (as opposed to background one), and subsequently compute the full expression for the number counts fluctuation. 
Section \ref{sec:gr_effects} explores the difference between the angular power spectra in redshift and luminosity distance spaces, as well as the relevance of
the various terms at different distances.
Finally, section \ref{sec:disc} is devoted to summary and conclusions. In Appendix \ref{sec:LDharm} we present a thorough calculation on how we computed the transfer functions implemented in \texttt{CAMB}, and a summary in \ref{sec:transfer}. Throughout this paper we will use natural units by fixing the speed of light to one, i.e., $c=1$.

\section{The observed number counts in luminosity distance space}\label{sec:contrast}

In this section we compute the number counts fluctuation for tracers in luminosity distance space, i.e. tracers for which we do not measure position or redshift but luminosity distance. First, we will compute the over-density in the number counts as a function of the perturbation in luminosity distance. We then add the effect of a flux-limited survey (magnification bias) for the two most relevant luminosity distance tracers, SNIa and GWs. Furthermore, we will identify the appropriate perturbation in luminosity distance to use in this derivation, and clearly lay out the computation of the volume perturbation when using luminosity distance as our observed quantity. Finally, in section \ref{sec:expression} we report the general expression including all relativistic corrections. Readers principally interested in the evaluation of our results can skip to section \ref{sec:expression} and then proceed to section \ref{sec:gr_effects}.

In this paper we will use a perturbed, flat FLRW metric in conformal Newtonian gauge given by
\be\label{eq:metric}
ds^2=a^2(\eta)\left[-(1+2\Psi)d\eta^2+(1-2\Phi)\delta_{ab}dx^a dx^b\right]\,,
\ee
where $\Phi$ and $\Psi$ are the metric potentials and $\eta$ is conformal time.
 
\subsection{Number density perturbation}
Let us consider a number of objects seen in a given (observed) direction $\bn$ at a given (observed) luminosity distance $D_L$ and denote it by $N(\bn,D_L)d\Omm_{\bn}dD_L$, where $\Omm_{\bn}$ is an infinitesimal solid angle element. \emph{Per se}, this quantity does not carry any relevant cosmological information. It is the variance of its fluctuation with respect to a cosmological average of the sources that one can use to relate to the primordial power spectrum. Therefore, we want to determine the fluctuation of this quantity at first order in perturbation theory.
To compute the number density contrast we then start from: 
\be
\Delta_N=\frac{N-\langle N\rangle}{\langle N\rangle}
\ee
where $\langle N\rangle$ is an average over directions. These quantities are determined at the observed luminosity distance, although ultimately we need to relate them to an affine parameter which one can take to be the background redshift. The number count is simply given by a number density times a volume, i.e., $N=\rho V$, then 
\be \label{eq:Delta_DL}
 \De(\bn,D_L) =\frac{\rho(\bn,D_L)V(\bn,D_L)-\bar\rho(D_L)\bar V(D_L)}{\bar\rho(D_L)\bar V(D_L)} \,,
\ee
where we used bars for average/background quantities. We now expand the observed volume and observed (physical, as opposed to comoving) number density around their background values, i.e., 
\begin{subequations}
\bea
\rho(\bn,D_L)&=&\bar\rho(D_L)+\de\rho(\bn,D_L) \,,\\
V(\bn,D_L) &=& \bar V(D_L)+\de V(\bn,D_L) \,.
\eea
\end{subequations}
Substituting these into \eqref{eq:Delta_DL}, one reaches
\be \label{eq:Delta_DL_retios}
 \De(\bn,D_L) =\frac{\de \rho(\bn,D_L)}{\bar\rho(D_L)}+\frac{\de V(\bn,D_L)}{\bar V(D_L)} \,.
\ee
This expression is similar to the one found in redshift space. However, we note that the end result will necessarily yield a different number count than that in redshift space. The size of an arbitrary cell, and thus the number of e.g. encompassed GW sources/SNIa events, will inherently depend on the observing quantity used (i.e. $z$ or $D_L$), affecting the first term on the RHS in eq. \eqref{eq:Delta_DL_retios}. Further, a perturbation in the latter won't be identical in both spaces, thus the volume (and the volume perturbation) of the cell will differ between redshift and luminosity distance space, altering the last term in eq. \eqref{eq:Delta_DL_retios}. Ultimately, the way the observed spaces relate to a background affine parameter is different.

So far we have expanded density and volume in terms of their background quantities but they are still written in terms of the observed luminosity distance. Therefore, we need to go from the observed $D_L$ to the background quantity, $\bar{D}_L$, through the affine parameter which we take to be the background redshift, $\bar{z}$; thus, $\bar D_L\equiv \bar D_L(\bar z)$. Note that in redshift space we perturb from observed to background quantities, and then from the observed parameter to the background parameter. Here we should do the same, i.e.,
\be\label{eq:lumdist_perturbed}
D_L=\bar D_L + \de D_L \,.
\ee
Note that in eq.\eqref{eq:Delta_DL_retios} one approximates $\bar V(D_L)\simeq \bar V(\bar D_L)$ as the numerator is already a perturbed quantity. The same is done for $\bar\rho(D_L)$. 

Next, we expand $\bar \rho$ to relate the density perturbation to a background quantity, i.e.:
\be
\bar \rho(D_L)= \bar \rho(\bar D_L + \de D_L)=\bar \rho(\bar D_L) + \frac{\partial \bar\rho}{\partial D_L}\bigg|_{D_L=\bar D_L}\de D_L\,.
\ee
Then,
\be
\de \rho(\bn,D_L)=\rho(\bn,D_L)-\bar\rho(D_L)=\rho(\bn, D_L)-\bar \rho(\bar D_L) - \frac{\partial \bar\rho}{\partial D_L}\bigg|_{D_L=\bar D_L}\de D_L\,,
\ee
therefore
\be
\frac{\de \rho(\bn,D_L)}{\bar\rho(D_L)}=
\de_n - \frac1{\bar\rho(\bar D_L)}\frac{\partial \bar\rho}{\partial D_L}\bigg|_{D_L=\bar D_L}\de D_L\, .\
\ee
where $\delta_n=\left(\rho(\bn,D_L)-\bar\rho (\bar D_L)\right)/\bar\rho (\bar D_L)$ is the density contrast in the Newtonian gauge.

Written in terms of the background redshift $\bar z$ and noting that the comoving number density $n$ is related to the physical one $\rho$ via $\rho=a^{-3}n$, then:
\bea
\frac{1}{\bar\rho}\frac{\partial \bar\rho}{\partial D_L}\bigg|_{D_L=\bar D_L} &=& \frac{1}{a^{-3}n}\frac{\partial (a^{-3}n)}{\partial D_L} = a^3n^{-1}\frac{\partial (a^{-3}n)}{\partial a}\left(\frac{\partial D_L}{\partial z}\frac{\partial z}{\partial a}\right)^{-1}\nn\\
%&=& a^3n^{-1}\left[-3a^{-4}n+a^{-3}\frac{\partial n}{\partial a}\right]\left(\frac{\partial D_L}{\partial z}\frac{\partial z}{\partial a}\right)^{-1} \nn\\
&=& \frac{1}{a}\left[-3+\frac{\partial\ln n}{\partial\ln a}\right]\bar{r}^{-1}\left(1+\frac{1}{\bar r\mathcal{H}}\right)^{-1}(-a^2) = \frac{\gamma}{\bar{D}_L}\left[3-\frac{\partial\ln n}{\partial\ln a}\right],\
\eea
where we used 
\bea
\frac{\d \bar D_L}{\d z}&=&\bar{r} + \frac1{\cal H}=\frac{\bar D_L}{(1+z)\gamma}\,,\\
\frac{\d a}{\d z} &=& -a^2 \, ,\\
\gamma &\equiv& \frac{\bar r \cal H}{1+\bar r \cal H}\,,
\eea
where $\bar r$ is the background comoving distance to the source. Using these results in eq. \eqref{eq:Delta_DL_retios} we have 
\be \label{eq:delta_}
 \De(\bn,D_L) =\de_n-\left[3-\frac{\partial\ln n}{\partial\ln a}\right]\gamma\frac{\de D_L}{\bar D_L}+\frac{\de V(\bn,D_L)}{\bar V(D_L)} \,,
\ee
where we note that the partial derivative of the comoving number density is the evolution bias \cite{Maartens_2021}.

In the case of a luminosity limited survey we need to include a further perturbation, which arises from lensing magnification. This can be thought of as the change in the observed number density of sources w.r.t. the change in the threshold of detection --- i.e. a perturbation in the observed distance might nudge an event above/below the minimum luminosity of the survey.  This term will depend on the type of tracer that populates our number density contrast, as it inherently depends on the means of observation. For gravitational waves, where $\sigma$ is the signal-to-noise ratio (SNR) of the GW in the detector network:
\bea
\Delta_O^{GW}(D_L,\hat{n},\sigma>\sigma_{th}) = \Delta_O^{GW}(D_L,\hat{n}) + \frac{\partial\ln n}{\partial\ln\sigma}\bigg|_{\sigma=\sigma_{th}}\times\frac{\delta\sigma}{\sigma},
\label{eq:abc}
\eea
where the first term on the RHS is the number density contrast for a survey not limited by luminosity/SNR, while the second is the magnification bias multiplied by a perturbation in the SNR.

In order to compute $\delta\sigma/\sigma$ we start from:
\bea
\sigma(D_L) = \sigma(\bar{D}_L+\delta D_L) = \sigma(\bar{D}_L)+\frac{\partial\sigma}{\partial D_L}\bigg|_{D_L=\bar{D}_L}\delta D_L,
\eea
which is equivalent to
\bea
\sigma(D_L) = \sigma(\bar{D}_L)+\delta\sigma.
\eea
From \cite{Finn_2000, Schutz_2011, Finn_1996, Oguri_2018} we find that the SNR is:
\be\label{eq:snr}
\sigma = \sqrt{\frac5{96\pi^{4/3}}} \frac1{D_L}(\mathcal{M}_z)^{5/6}\theta\sqrt{I} \propto \frac1{D_L},
\ee
where $I$ is an integral over frequency which includes the sensitivity curve of the detector\footnote{In the case of a network of detectors the signal-to-noise is added in quadrature, retaining the proportionality to $D_L^{-1}$.}, $\theta$ is a function of angles that describes the orientation of the binary with respect to the detector and $\mathcal{M}_z$ the redshifted chirp mass.
Thus, using eq. \eqref{eq:lumdist_perturbed}:
\bea
\frac{\delta\sigma}{\sigma(\bar{D}_L)} &=& \frac{\sigma(D_L)-\sigma(\bar{D}_L)}{\sigma(\bar{D}_L)} = \frac{\bar{D}_L}{D_L}-1 = \left(1+\frac{\delta D_L}{\bar{D}_L}\right)^{-1}-1 \simeq -\frac{\delta D_L}{\bar {D}_L}\,,
\eea
where we used binomial expansion in the last step.
Therefore
\bea
\frac{\delta\sigma}{\sigma} \simeq -\frac{\delta D_L}{\bar{D}_L}\,,
\eea
where the relative minus sign implies that an increase in the observed luminosity distance decreases the corresponding SNR, as it should.

We can now rewrite the number density contrast from eq. \eqref{eq:abc} as:
\bea\label{eq:dens_contrast_bias_GW}
\Delta_O^{GW}(D_L,\hat{n},\sigma>\sigma_{th}) = \de_n-\left[3-\frac{\partial\ln n}{\partial\ln a}+\frac{1}{\gamma}\frac{\partial\ln n}{\partial\ln\sigma}\right]\gamma\frac{\de D_L}{\bar{D}_L}+\frac{\de V(\bn,D_L)}{\bar V(D_L)} \,.
\eea
Again, we stress that the partial derivatives w.r.t. the scale factor and the SNR are, respectively, the evolution and magnification bias for GWs.
However, eq. \eqref{eq:dens_contrast_bias_GW} is specific to GWs because of the bias term unique to this case. Instead, in the case of Supernovae Type Ia, we note that the survey will be {apparent} magnitude {$m$} limited:
\be
\Delta_O^{SNIa}(D_L,\hat{n},{m<m_{max}}) = \Delta_O^{SNIa}(D_L,\hat{n}) {+ \frac{\partial\ln n}{\partial\ln m}\bigg|_{m=m_{max}}\times\frac{\delta m}{m}}+\frac{\de V(\bn,D_L)}{\bar V(D_L)}.
\ee
Recalling the definition of absolute magnitude $M$
\bea
M = m - 5\log_{10}(D_L[\mathrm{Mpc}])+25 \,,
\eea
%where $m$ is the apparent magnitude, 
it is then straightforward to calculate the {$\delta m$} term, {as $M$ is fixed for all SNIa}. Combining it with the evolution bias term (unchanged) we find:
\be\label{eq:dens_contrast_bias_SN}
\Delta_O^{SNIa}(D_L,\hat{n},{m<m_{max}}) = \delta_n - \left[3 - \frac{\partial\ln n}{\partial\ln a} {-}5\frac{1}{\gamma}\frac{\partial\log_{10}n}{\partial {m}}\right]\gamma\frac{\delta D_L}{\bar{D}_L} +\frac{\de V(\bn,D_L)}{\bar V(D_L)} .
\ee
For simplicity, let us define here the evolution and magnification bias terms:
\begin{subequations}
\bea
b_e = \frac{\partial\ln n}{\partial\ln a}
\eea
\be
s = 
\begin{dcases*}
    {-}\frac1{5}\frac{\partial\ln n}{\partial\ln\sigma} ,& for GWs;\\
    \frac{\partial\log_{10} n}{\partial {m}} ,& for SNIa\\
\end{dcases*}
\ee
\end{subequations}
{Note that the magnification biases for GWs and SNIa have opposite signs. In fact, whilst a higher signal-to-noise ratio threshold will reduce the amount of sources seen, a higher magnitude threshold will increase it, as the two observables behave oppositely.}

Thus we can write the general case:
\bea \label{eq:general_num_contrast}
\De_O(D_L,\hat{n}) = \de_n-\left[3-b_e{-}\frac{5}{\gamma}s\right]\gamma\frac{\de D_L}{\bar{D}_L}+\frac{\de V(\bn,D_L)}{\bar V(D_L)} \, .
\eea
To proceed we therefore need to compute the luminosity distance perturbation with respect to a background affine parameter, as well as the volume perturbation.

\subsection{Luminosity distance perturbation}
To evaluate eq. \eqref{eq:general_num_contrast} we now need to specify the appropriate perturbation in luminosity distance, $\delta D_L/{\bar{D}_L}$. %; for this we require an expression for an observed quantity in terms of perturbed background. 
{Therefore, we redo the calculation to obtain the expression for the perturbation of the luminosity distance at a background affine parameter in terms of the perturbed quantities relevant for cosmology, and which are commonly used in publicly available codes. Alternative calculations of $\delta D_L$ in a perturbed FLRW universe can be found in \cite{Sasaki:1987ad,Pyne:2003bn,Hui:2005nm}. Note that other authors have computed the luminosity distance perturbation at the observed redshift \cite{Bertacca2018,Bonvin:2005ps}, leading to a slightly different expression, as expected. For the sources we consider in this paper we do not have access to an observed redshift. Our purposes here is to determine the luminosity distance perturbation with respect to the background luminosity distance at a background affine parameter. This perturbation will be valid not only for transients where we directly measure the luminosity distance, but also any stochastic gravitational background.} 

We start from the distance-duality relationship between the luminosity distance and the area distance for an observed source, which is valid as long as geometric optics holds and photon number is conserved \cite{Etherington,Fleury_2015}, and is given by:
\be \label{eq:lumdist_start}
D_L(z_s)=(1+z_s)^2 D_A(z_s) \, .\
\ee
From \cite{Bacon2014}, who consider the area distance, we have:
\bea\label{eq:da_bacon}
D_{A}(z_{s})&=&\bar{D}_{A}(z_{s})(1-\kappa(\bm n))%\nn\\
%&=&\bar{D}_{A}(z_{s})\left\{1+\frac{\delta D_{A}}{\bar{D}_{A}}\bigg|_{z_{s}}+\left[1-\frac{1+z_{s}}{H\left(z_{s}\right) r_{s}\left(z_{s}\right)}\right] \frac{\delta z}{1+\bar{z}}\bigg|_{z_{s}}\right\}
\eea
which is the \emph{area distance $D_A$ to the source at observed redshift $z_s$} at direction $\bm n$. Note that barred objects refer to background quantities. 
The perturbation at the observed redshift which appears in eq. \eqref{eq:da_bacon} is then found in \cite{Bertacca2018,Bacon2014}:
\bea \label{eq:kappa}
\kappa(\boldsymbol{n}) &=&-\left(1-\frac1{\bar{r}\mathcal{H}}\right)\boldsymbol{v}\cdot\boldsymbol{n}+\frac1{2}\int_{0}^{\bar r} \d r\ \frac{\bar{r}-r}{\bar{r}r}\Delta_\Omm(\Phi+\Psi) -\frac{\Psi}{\bar r\mathcal{H}}+\left(1-\frac1{\bar r\mathcal{H}}\right)\int_0^{\bar r}\d r\ (\Phi'+\Psi') \nn\\
&&+ (\Phi+\Psi) -\frac1{\bar r}\int_{0}^{\bar r}\d r\ (\Phi+\Psi)
\eea
Note that at the ``observed'' redshift, multiplying both sides of eq. \eqref{eq:da_bacon} by $(1+z_s)^2$ implies
\bea \label{eq:dlpertzs}
D_{L}(z_{s})&=&\bar{D}_{L}(z_{s})(1-\kappa(\bm n)) \, ,
\eea
meaning that at the observed redshift the perturbation to the luminosity distance is the same as for the area distance. This result is widely known and agrees with the literature \cite{Bertacca2018,Bonvin:2005ps}.

However, to compute the over-density of tracers, we need instead the perturbation as a function of background affine parameter distance, which we take to be the \textit{background redshift}. In this analysis we do not observe a redshift, only an observed luminosity distance. In order to relate it to background quantities we exploit the fact that an observed $D_L$ has a corresponding $z_s$ which is relatable to the background $\bar z$. The latter is clearly unobservable, but it is required to describe the background distribution of sources.

Expanding $z_s=\bar{z}+\delta z$, the luminosity distance from eq. \eqref{eq:lumdist_start} written in terms of background quantities is then
\begin{align}
D_L(z_s,\boldsymbol{n})&=(1+\bar z+\delta z)^2\, \bar{D}_A(\bar z+\delta z)\left[1-\kappa(\boldsymbol{n})\right]\nn\\
&= \left[(1+\bar z)^2+2\delta z(1+\bar z)\right]\bar{D}_A(\bar{z})\left[1-\kappa(\boldsymbol{n})+\frac1{\bar{D}_A(\bar{z})}\frac{\d D_A}{\d z}\bigg|_{z_s}\delta z\right]\nn\\
&=\bar D_L(\bar z)\left[1-\kappa(\boldsymbol{n})+\frac{2\delta z}{1+\bar z}+\frac1{\bar{D}_A(\bar{z})}\frac{\d D_A}{\d z}\bigg|_{z_s}\delta z\right]\nn\\
&= \bar D_L(\bar z)\left[1-\kappa(\boldsymbol{n})+\left(1+\frac1{\mathcal{H}\bar r}\right) \frac{\delta z}{1+\bar{z}}\right] \,.
\end{align}
Let us \textit{define} the luminosity distance perturbation with respect to a background affine parameter in the same manner as done in eq. \eqref{eq:dlpertzs}, i.e. 
\be
{\delta D}_L\equiv  D_L(z_s,\boldsymbol{n})- \bar D_L(\bar z)\,.
%=-\tilde\kappa(\boldsymbol{n})\bar D_L(\bar z)\,.
\ee
%Notice the tilde to differentiate this perturbation from $\kappa$. In this paper, $\tilde\kappa$ is the luminosity distance perturbation which we will always refer to. 
Hence, the luminosity distance fluctuation we need is
\be \label{eq:lumdistperturb3}
\frac{\delta D_L}{\bar D_L}(\bar z)=-\kappa(\boldsymbol{n})+\left(1+\frac{1}{{\mathcal H} \bar r}\right) \frac{\delta z}{1+\bar{z}}.
\ee
To linear order we have
\bea\label{eq:dz}
\frac{\delta z}{1+\bar{z}} = \boldsymbol{v}\cdot\boldsymbol{n}-\Psi-\int_0^{\bar r} \d r\ (\Phi'+\Psi'),
\eea
and combining it with eq. \eqref{eq:kappa} in \eqref{eq:lumdistperturb3} gives
\be\label{eq:epsilon}
\frac{\delta D_L}{\bar D_L} = 2\boldsymbol{v}\cdot\boldsymbol{n} -\frac1{2}\int_{0}^{\bar r} \d r\ \frac{\bar r-r}{\bar r r}
\Delta_\Omm(\Phi+\Psi) -\Phi-2\Psi-2\int_0^{\bar r}\d r\ (\Phi'+\Psi')+\frac1{\bar r}\int_0^{\bar r}\d r\ (\Phi+\Psi) \, ,
\ee
which agrees with the perturbation identified in \cite{Namikawa}{, and already computed in \cite{Sasaki:1987ad,Pyne:2003bn,Hui:2005nm}}. Note that our derivation of the luminosity distance perturbation at the background affine parameter is different from that found in \cite{Namikawa}, but yields the same result. {Our derivation is in fact similar to \cite{Pyne:2003bn,Hui:2005nm}, using as starting point the distance-duality relationship.}

%%%%%%%%%%
\subsection{Volume perturbation}
 
The final part of our calculation of the number count over-density is to determine the volume perturbation in eq. \eqref{eq:general_num_contrast}.
%We start by defining our FLRW metric in conformal Newtonian Gauge:
%\bea\label{eq:metric}
%ds^2=a^2(\eta)\left[-(1+2\Psi)d\eta^2+(1-2\Phi)\delta_{ab}dx^a dx^b\right]
%\eea
%
To do so we follow closely \cite{Bonvin2011}. Note that we will use similar notation but we write quantities in terms of luminosity distance. First let us start with the volume element:
\bea
\d V &=& \sqrt{-g}\ep_{\mu\nu\al\beta}\,u^\mu \d x^\nu\,\d x^\al\,\d x^\beta = 
\sqrt{-g}\;\epsilon_{\mu\nu\al\beta}\,u^\mu\! \frac{\dd x^\nu}{\dd D_L} \! 
\frac{\dd x^\al}{\dd \theta_s}\! \frac{\dd x^\beta}{\dd \varphi_s}\! 
\left| \frac{\dd (\theta_s,\varphi_s)}{\dd (\theta_o,\varphi_o)} 
\right|\,\d D_L\,\d \theta_o\,\d\varphi_o \nonumber \\  \label{eN:vol}
&\equiv& v(D_L,\theta_o,\varphi_o)\,\d D_L\,\d \Omm_o ~,
\eea
where $\d\Omm_o=\sin\theta_o\d\theta_o\d\varphi_o$, and $\epsilon_{\mu\nu\al\beta}$ is the Levi-Civita symbol. Given that the volume density in terms of the observer's coordinates is defined as
\be
v\equiv \sqrt{-g}\;\epsilon_{\mu\nu\al\beta}u^\mu\! \frac{\dd x^\nu}{\dd D_L} \! 
\frac{\dd x^\al}{\dd \theta_s}\! \frac{\dd x^\beta}{\dd \varphi_s}\! 
\left| \frac{\dd (\theta_s,\varphi_s)}{\dd (\theta_o,\varphi_o)} 
\right|\,,
\ee
we can compute the volume perturbation as
\be
\frac{\de V}{V} =   \frac{v -\bar v}{\bar v} =  \frac{\de v}{\bar v}\,.
\ee
%
%The volume density is defined as  
%\be
%v\equiv \sqrt{-g}\;\epsilon_{\mu\nu\al\beta}u^\mu\! \frac{\dd x^\nu}{\dd D_L} \! 
%\frac{\dd x^\al}{\dd \theta_s}\! \frac{\dd x^\beta}{\dd \varphi_s}\! 
%\left| \frac{\dd (\theta_s,\varphi_s)}{\dd (\theta_o,\varphi_o)} 
%\right|\,,
%\ee
%in terms of the observer {\color{red}(be more precise -- in terms of observer's coordinates or something? Also, to me it seems preferable to define $v$ via 2.38 before using it in 2.37, but not a major issue if you prefer this way round)}. 
Recalling that we need to relate all quantities to background parameters (e.q. formally defining $\bar{r}$ via $\bar{D_L}$), one should note that the background volume density in luminosity distance space is given by
\be 
\bar{v}(\bar D_L)=\frac{a^4\bar{r}^2(\bar{D}_L)}{1+\mathcal{H}\bar{r}(\bar{D}_L)}\,.
\ee
The angles at the source and the observed angles are related via a small perturbation:
\bea 
\theta_s&=&\theta_0+\de \theta \label{eq:thetapert}\, ,\\
\varphi_s&=&\varphi_0+\de \varphi \label{eq:varphipert}\,.
\eea
Then, to first order the determinant of the Jacobian of the transformation of the angular position is
\be
\left| \frac{\dd (\theta_s,\varphi_s)}{\dd (\theta_o,\varphi_o)} 
\right|=1+\frac{\dd \de \theta}{\dd \theta}+\frac{\dd \de \varphi}{\dd \varphi}\,.
\ee 
Together with the fact that the determinant of the metric is $\sqrt{-g}=a^4(1+\Psi-3\Phi)$, and that the 4-velocity of the source is $u=(1-\Psi,v^i)$, one finds that the perturbed volume density can be expressed as
\be \label{eq:pert_v}
v(D_L)=a^3(1+\Psi-3\Phi)\Bigg[ \frac{\d r}{\d D_L} r^2\frac{\sin\theta_s}{\sin\theta_o} 
\left(1+\frac{\dd \de \theta}{\dd \theta}+\frac{\dd \de \varphi}{\dd \varphi}\right)-\left(\Psi\frac{\d \bar r}{\d \bar D_L}+\boldsymbol{v}\cdot\boldsymbol{n}\frac{\d \bar \eta}{\d \bar D_L}\right) \bar r^2\Bigg] .
\ee
Using the angle perturbation in eq. \eqref{eq:thetapert} we have
\be \label{eq:sine}
\sin{\theta_s}\simeq \sin{\theta_0}+\cos{\theta_0}\ \de \theta = \sin{\theta_0} \left(1+ \cot{\theta_0}\ \de \theta\right)\,.
\ee 
One should also note that the comoving distance is a perturbed quantity, i.e, $r^2=(\bar r+\de r)^2\simeq \bar r^2 +2 \bar r \de r$. 
%In the last terms of \eqref{eq:pert_v} all derivatives are in terms of background quantities as they are already multiplied by first order quantities (note the two terms in square bracket different - one background one not) {\color{red}(almost true -- but isn't there a leading order term in there right now? [It will get subtracted off later\ldots])}. 
We then have
\be\label{eq:drDl}
\frac{\d \bar r}{\d \bar D_L}=\left(1+\bar{z} +H\bar{r} \right)^{-1}=\frac1{(1+\bar{z})(1+\bar{r}\cal H)}\,,
\ee
and
\be\label{eq:detaDl}
\frac{\d \bar \eta}{\d \bar D_L}=-\frac{\d \bar r}{\d \bar D_L}=-\frac1{(1+\bar z)(1+\bar{r}\cal H)}\,.
\ee
We can write at linear order
\be\label{eq:derrDL}
\frac{\d r}{\d D_L}=\frac{\d\bar r}{\d\bar D_L}+\frac{\d \de r}{\d\bar D_L}-
\frac{\d \de D_L}{\d\bar D_L}\frac{\d \bar r}{\d \bar D_L}=\left(\frac{\d \bar r}{\d\eta}+
\frac{\d \de r}{\d\la}-\frac{\d\de D_L}{\d\la}\frac{\d\bar r}{\d \bar D_L} \right)\frac{\d\eta}{\d\bar D_L} ,
\ee
where we take derivatives along the photon geodesic and set $\d \eta=\d \lambda$. Note as well that $\d \bar r /\d\eta=-1$. Then, expanding eq. \eqref{eq:derrDL} with eqs. \eqref{eq:drDl} and \eqref{eq:detaDl}, and substituting it into eq. \eqref{eq:pert_v}, together with the expansion described in eq. \eqref{eq:sine}, we obtain:
\be 
v(D_L)=\frac{a^4\bar r^2}{1+\bar r\cal H}\left[1-3\Phi-\frac{\d \de r}{\d\la}+\frac{a}{1+\bar r\cal H}\frac{\d\de D_L}{\d\la}+\frac{2\de r}{\bar r}+\left( \cot\theta + \frac{\dd}{\dd \theta}\right)\de \theta +\frac{\dd \de \varphi}{\dd \varphi}+\boldsymbol{v}\cdot\boldsymbol{n}\right]\,.
\ee 
Finally we need to perturb $v$ around the background $D_L$, i.e. 
\be 
\bar v(D_L) = \bar v(\bar D_L) + \frac{d\bar v}{d\bar D_L}\de D_L\,.
\ee
where
\be 
\frac{d\bar v}{d\bar D_L}=\frac{a^4\bar r^2}{1+\bar r\cal H}\left[\frac2{\bar r \cal H}+\frac1{1+\bar r\mathcal{H}}\left(\bar r\mathcal{H}\frac{{\cal H}'}{{\cal H}^2}-1\right)-4\right]\frac{\cal H}{(1+z)(1+\bar r\mathcal{H})}\,.
\ee 
Then the volume perturbation is given by
\bea \label{eq:volpert2}
\frac{\de v}{\bar v}(\bn,D_L) &=& \frac{v(D_L,\bm{n}) -\bar v(D_L)}{\bar v(D_L)}\nn\\ 
&=& -3\Phi-\frac{\d \de r}{\d\la}+\frac{a}{1+\bar r\cal H}\frac{\d\de D_L}{\d\la}+\frac{2\de r}{\bar r}+\left( \cot\theta + \frac{\dd}{\dd \theta}\right)\de \theta +\frac{\dd \de \varphi}{\dd \varphi}+\boldsymbol{v}\cdot\boldsymbol{n}\nonumber\\ 
&&- \left[\frac2{\bar r \cal H}+\gamma\left(\frac{\cal{H}'}{\mathcal{H}^2}-\frac1{r\cal{H}}\right)-4\right]\frac{{\cal H} \de D_L}{(1+z)(1+\bar r\mathcal{H})}\, .
\eea

Using previous results shown in \cite{Bonvin2011} one can start expressing the volume perturbation in terms of perturbed quantities such as the potentials and peculiar velocity. The perturbation in position is given by
\be 
\de r =\int_{0}^{\bar r}\d r (\Phi+\Psi) \,. 
\ee 
Then 
\be\label{eq:drdl}
\frac{\d \de r}{\d \lambda} =-(\Phi+\Psi)\,,
\ee 
and 
\be\label{eq:drr}
2\frac{\de r}{\bar r} =\frac2{\bar r}\int_{0}^{\bar r}\d r\ (\Phi+\Psi) \,, 
\ee 
is the standard Sachs-Wolfe term. 
%At this point we also note that to lowest order $r_s \approx \bar r$, and hence substitute accordingly.
The angular perturbations give rise to the traditional kappa term caused by lensing
\be\label{eq:kappag}
\left( \cot\theta + \frac{\dd}{\dd \theta}\right)\de \theta +\frac{\dd \de \varphi}{\dd \varphi}=-2\kappa_g=-\int_{0}^{\bar r} \d r\ \frac{\bar r-r}{\bar r r}
\Delta_\Omm(\Phi+\Psi)\,.
\ee
Further, expressing $\delta D_L = \frac{\delta D_L}{\bar{D}_L}\,\bar{D}_L$, we can write
\bea\label{dDldl}
\frac{\d\de D_L}{\d\la}=\bar{D}_L\frac{\d}{\d\lambda}\left(\frac{\delta D_L}{\bar{D}_L}\right)+\frac{\delta D_L}{\bar{D}_L}\frac{\d D_L}{\d\lambda}\,.
\eea
Finally, substituting eqs. \eqref{eq:drdl},\eqref{eq:drr},\eqref{eq:kappag} and \eqref{dDldl} into eq. \eqref{eq:volpert2} yields:
\bea \label{eq:volpert_midcalc}
\frac{\de v}{\bar v}(\bn,D_L) &=& -2\Phi+\Psi +\frac{\gamma}{\cal H}\frac{\d}{\d \lambda}\left(\frac{\delta D_L}{\bar{D}_L}\right)+\boldsymbol{v}\cdot\boldsymbol{n}-2\kappa_g+\frac2{\bar r}\int_{0}^{\bar r}\d r\ (\Phi+\Psi) \nn\\
&&+\frac{\delta D_L}{\bar{D}_L}\left(\vnd{-1}
-\gamma\left[\frac2{\bar r \cal H}+\gamma\left(\frac{\cal{H}'}{\mathcal{H}^2}-\frac1{\bar r\cal{H}}\right)-4\right]\right)\,.
\eea
We need to calculate the derivative with respect to the affine parameter along the photon geodesic, which we take to be the conformal time when considering the total derivative. We also take
\be 
\frac{\d}{\d \lambda}=\frac{\d}{\d \eta}=\frac{\partial}{\partial \eta}-\boldsymbol{n}^i\frac{\partial}{ \partial \boldsymbol{x}^i}\,.
\ee
Then
\bea\label{eq:dk}
\frac{\d}{\d\lambda}\frac{\delta D_L}{\bar{D}_L}= -2(\mathcal{H}+\partial_r)\boldsymbol{v}\cdot\boldsymbol{n}+\Phi'+\partial_r\Phi-\frac{\Phi+\Psi}{r_s}+\frac1{2\bar r^2}\int_0^{\bar r} \d r\ \left[2+ \Delta_\Omega\right](\Phi+\Psi)\, ,\
\eea
where we also used the Euler Equation, i.e, $v_r'+{\cal H}v_r+\partial_r\Psi=0$. 

Finally, we obtain the expression for the volume perturbation in luminosity distance space:
\bea\label{eq:final_volume}
\frac{\de v}{\bar v}(\bn,D_L) &=& -2\Phi+\Psi +\frac{\gamma}{\cal H}\left[-2(\mathcal{H}+\partial_r)\boldsymbol{v}\cdot\boldsymbol{n}+\Phi'+\partial_r\Phi-\frac{\Phi+\Psi}{r_s} +\frac1{2\bar r^2}\int_0^{\bar r} \d r\ \left[2+ \Delta_\Omega\right](\Phi+\Psi)\right] \nn\\
&&+\boldsymbol{v}\cdot\boldsymbol{n}-2\kappa_g+\frac2{\bar r}\int_{0}^{\bar r}\d r\ (\Phi+\Psi) 
+\frac{\delta D_L}{\bar{D}_L}\left(\vnd{-1}
-\gamma\left[\frac2{\bar r \cal H}+\gamma\left(\frac{\cal{H}'}{\mathcal{H}^2}-\frac1{\bar r\cal{H}}\right)-4\right]\right) \, .\
\eea

\subsection{The observed density contrast in luminosity distance space}\label{sec:expression}

Now that we have all the components, we can expand eq. \eqref{eq:general_num_contrast} using eqs. \eqref{eq:epsilon}, \eqref{eq:final_volume} to find:
\bea\label{eq:num_final}
\De(\bn,D_L) &=&\de_n-2\Phi+\Psi+\frac{\gamma}{\mathcal{H}}\frac{\d}{\d\lambda}\frac{\delta D_L}{\bar{D}_L}+v_r-2\kappa_g+\frac{2}{\bar r}\int_0^{\bar r} \d r\ (\Phi+  \Psi) -\beta\frac{\delta D_L}{\bar{D}_L} \, \\
&=&\de_n+A_D(\bm v\cdot\bm n)+A_{LSD}\partial_r(\bm v\cdot\bm n) +A_{\Psi}\Psi + A_{\Phi}\Phi+ A_{\Phi'}\Phi'+A_{\nabla\Phi}\partial_r\Phi\nn\\
&&+\frac1{\bar r}\int_0^{\bar r} \d r\ (A_{TD}+A_{L}\Delta_{\Omega})(\Phi+\Psi)+A_{ISW}\int_0^{\bar r} \d r\ (\Phi'+\Psi')\label{eq:finalAs} \, ,\
\eea
recalling that $\gamma \equiv \frac{\bar{r}\mathcal{H}}{1+\bar{r}\mathcal{H}}$, and where the coefficient of the perturbation in $D_L$-space is:
\bea
\beta&\equiv&
\gamma\left[\frac2{\bar r \mathcal{H}}+\gamma\left(\frac{\cal{H}'}{\mathcal{H}^2}-\frac1{\bar r\cal{H}}\right)-1-b_e\right] {-}5s
\vnd{+1}\, ,
\eea
and where we grouped the terms of the various distortion effects into the following coefficients:
\begin{subequations}
\bea%\label{eq:terms}
&A_D& = 1-2(\gamma+\beta) \label{eq:AD}\, ,\\
&A_{LSD}& = -2\ \frac{\gamma}{\mathcal{H}} \label{eq:ALSD}\, ,\\
&A_{\Psi}& = 1-\frac{1}{1+\bar r\mathcal{H}}+2\beta \,,\\
&A_{\Phi}& = -2-\frac{1}{1+\bar r\mathcal{H}}+\beta \,,\\
&A_{\Phi'}& = \frac{\gamma}{\mathcal{H}} \, ,\\
&A_{\nabla\Phi}& = \frac{\gamma}{\mathcal{H}} \,,\\
&A_{TD}& = 2-\beta+ \frac{1}{1+\bar r \mathcal{H}} \,,\\
&A_{L}& = \frac12 \left[\left(\frac{\bar r-r}{r}\right)(\beta-2)+\frac{1}{1+\bar r \mathcal{H}}\right] \label{eq:AL} \, ,\\
&A_{ISW}& = 2\beta \, .
\eea 
\end{subequations}
These coefficients belong to, respectively, density, gradient of velocity (luminosity distance space distortions --- LSD), potentials and derivatives of potentials, time-delay (Sachs-Wolfe), lensing and ISW. The expression is more complex than the corresponding one in redshift space, and a comparison of the terms can be found in Appendix \ref{sec:table}. In particular, the latter is missing the term proportional to the gradient of the potentials. The magnification and evolution biases enter in most terms (with the exception of the derivatives of potentials and the LSD term); however, for the remainder of the paper these are neglected, as their modelling is left for future work. {One should note that $A_{LSD}$ had already been identified and included in the calculation of the angular power spectrum \cite{Libanore2021,Libanore2022}, and the 3D power spectrum and its multipoles \cite{Namikawa,Zhang:2018nea}. The lensing term $A_L$, with \textit{kappa} $\kappa_g$ and its line-of-sight gradient, is included in \cite{Libanore2022,Namikawa}, although their equivalent $\beta$ coefficient has no magnification bias included.}

The number density contrast $\delta_n$ is in the Newtonian gauge, but the bias is implicitly defined in the synchronous gauge with respect to the matter density contrast. If the number density of sources evolves with time then the density contrast in the two gauges is related via the expression \cite{Challinor_Lewis,Wands:2009ex}
\be\label{eq:gauge}
\delta_n = b\delta_M^{syn}+ \left[b_e-3\right]\frac{\mathcal{H} v}{k}\, ,\
\ee
where $\delta_M^{syn}$ is the matter density contrast in the synchronous gauge and $b$ is the tracer bias.

Note that hereafter we use redshift and luminosity distance interchangeably as a measure of distance. Although we will compute the angular power spectrum in luminosity distance, redshift is the conventional measure used to locate the line-of-sight distance to a bin of tracers and its width. One has a better intuition of distance using $z$ rather than $D_L$. At the background level one can convert between the two if we fix the cosmology. However, at the perturbed level, the observed and background quantities are related differently with the metric and energy-momentum tensor perturbations, as we have shown here and is part of the point of our paper.

\begin{figure}
    \centering
    \includegraphics[width=0.48\textwidth]{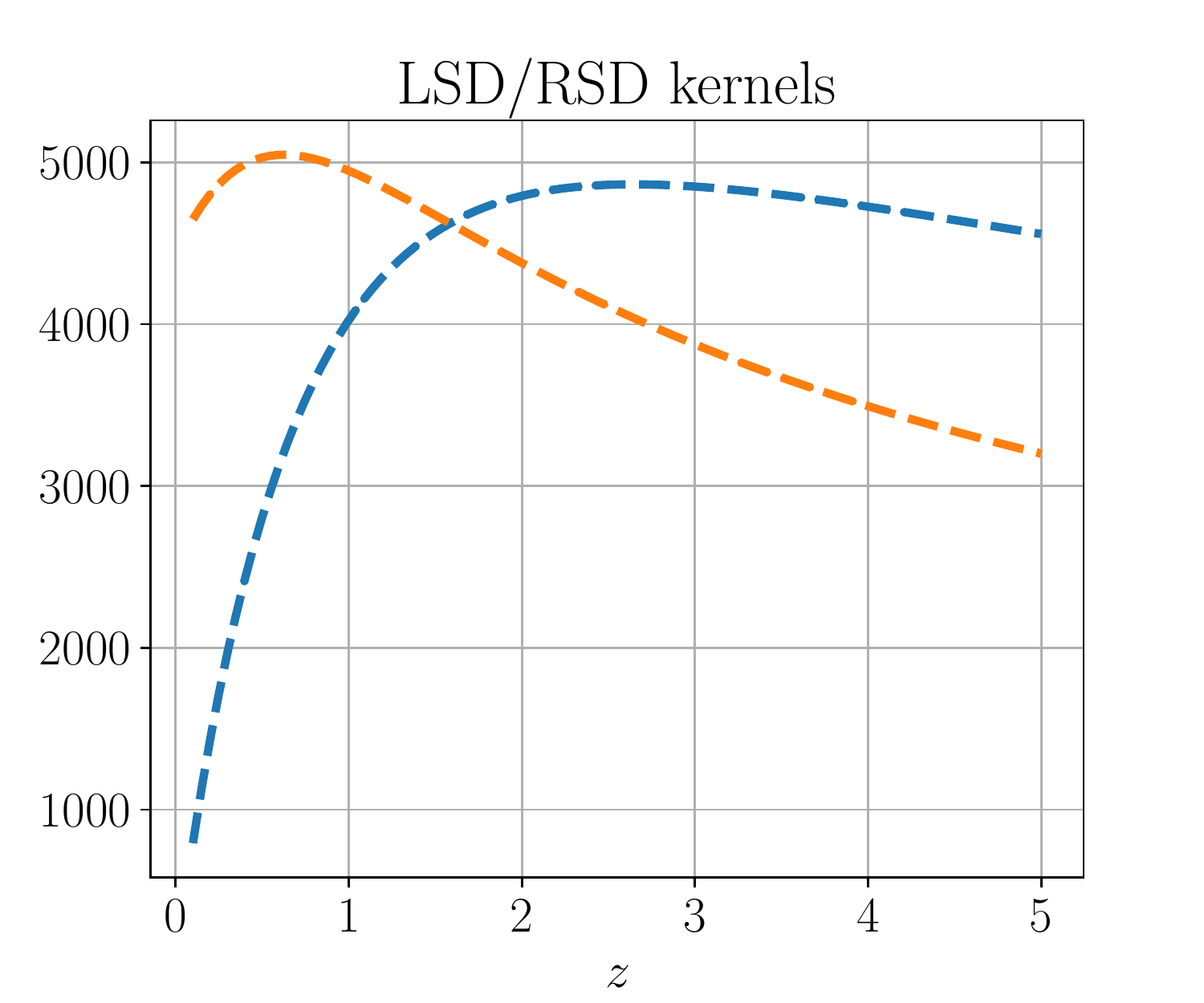}
    \includegraphics[width=0.48\textwidth]{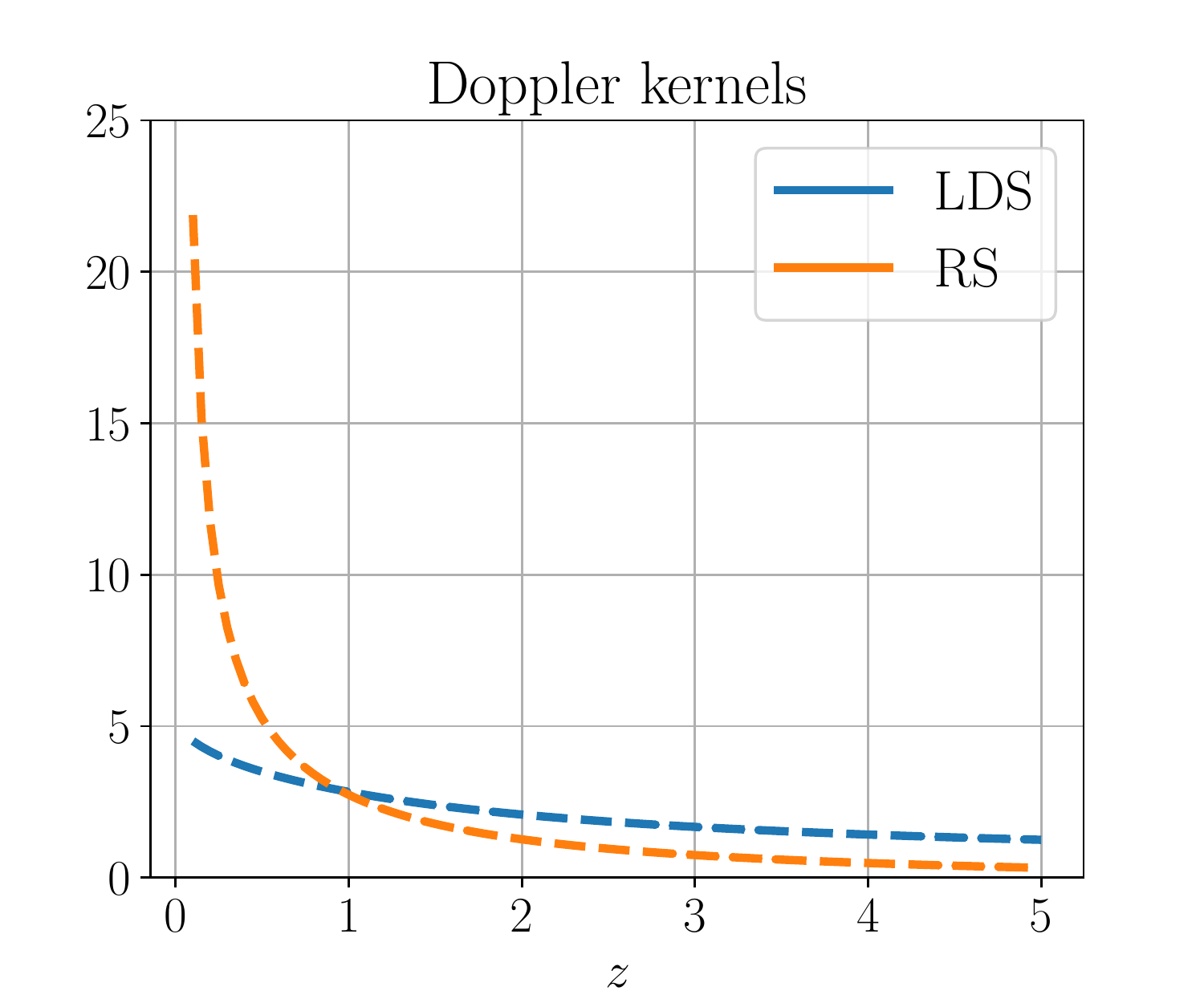}
    \caption{Comparison between kernels in luminosity distance space (blue) and redshift space (orange) as a function of redshift. \emph{Left}: LSD/RSD;
    \emph{Right:} Doppler. Dashed lines indicate negative values. We remind the reader that these are kernel amplitudes, and therefore dimensionless. Further, we set the magnification and evolution biases, $s$ and $b_e$ to zero to purely investigate the kernel.}
    \label{fig:kernelsV}
\end{figure}

\begin{figure}
    \centering
    \includegraphics[width=0.48\textwidth]{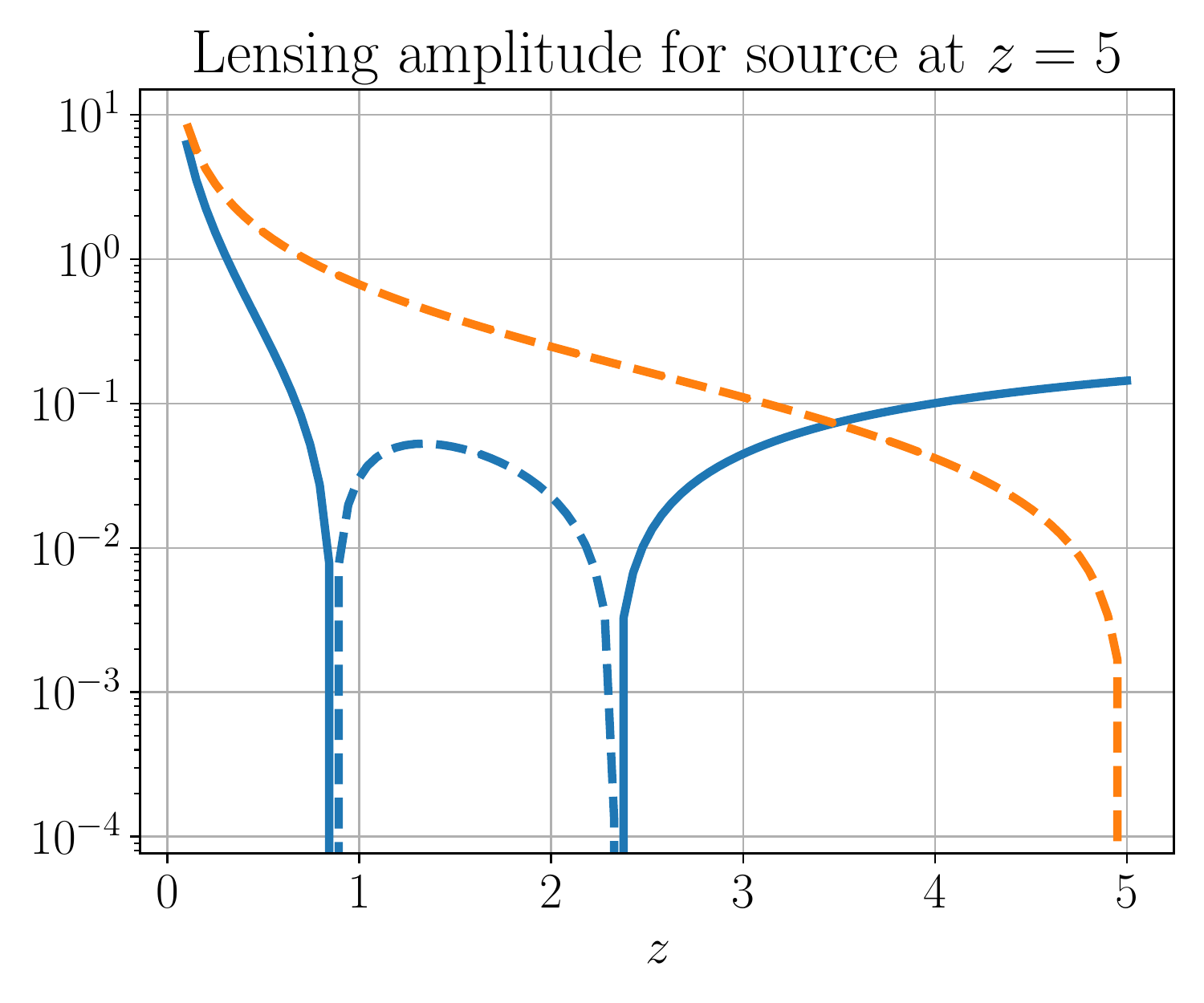}
    \includegraphics[width=0.48\textwidth]{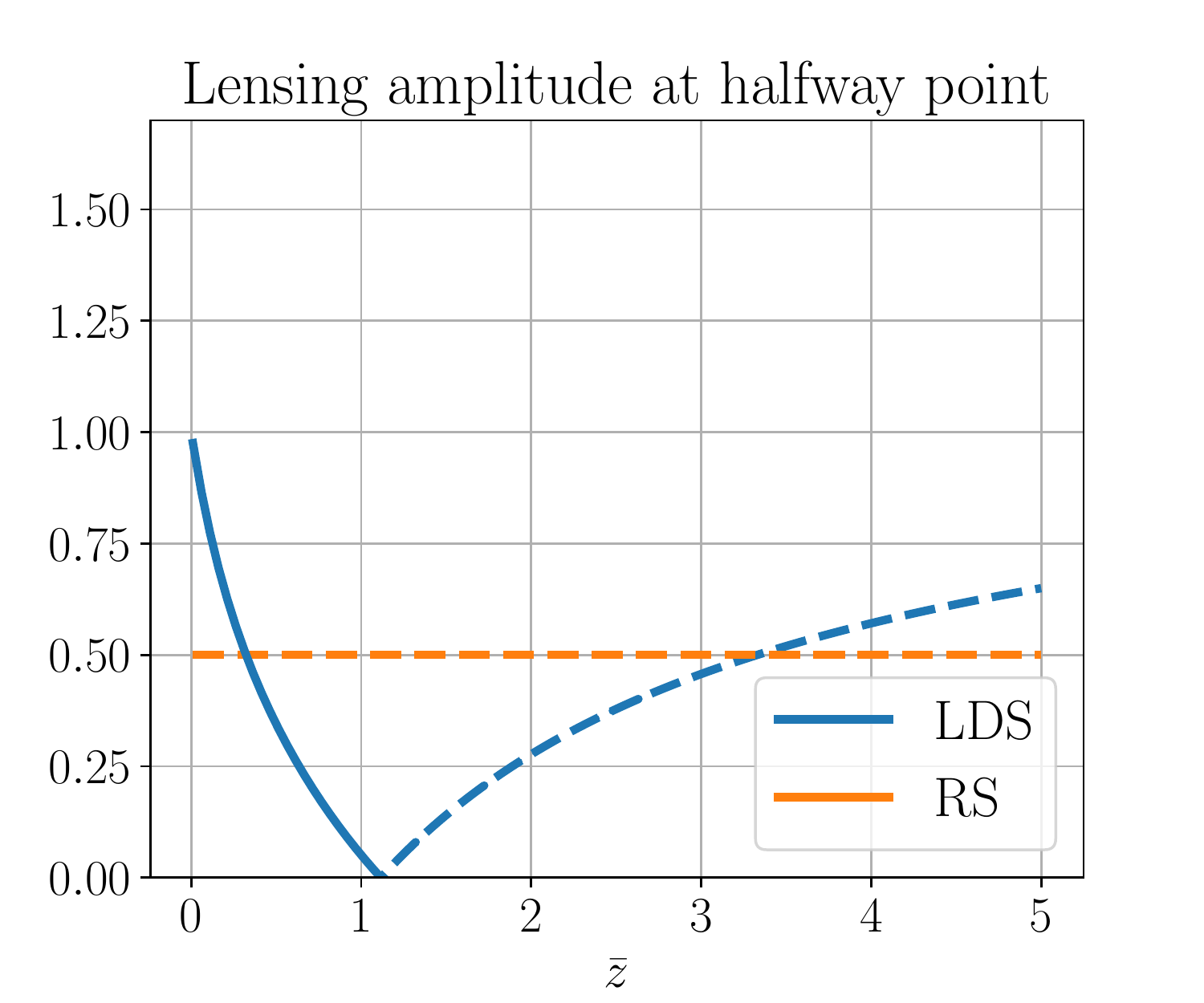}
    \caption{
    Comparison between $A_L$ in luminosity distance space (blue) and redshift space (orange). \emph{Left}: as a function of $r(z)$; \emph{Right:} as a function of $\bar r (z)$ (i.e. source distance) at a fixed $r=0.5\bar r$. Dashed lines indicate negative values. The magnification and evolution biases, $s$ and $b_e$, were set to zero.}
    \label{fig:kernelsL}
\end{figure}

The underlying matter density contrast in Newtonian gauge $\delta_n$ will be the same in both redshift and luminosity distance space, as expected.
However, the correction terms from \eqref{eq:num_final} differ significantly from the RS case. Whilst we investigate the effects of some important terms in this section and in section \ref{sec:gr_effects}, an interested reader can find the full terms comparison in appendix \ref{sec:table} (see as well \cite{Challinor_Lewis, DiDio_2013}). Figure \ref{fig:kernelsV} and figure \ref{fig:kernelsL} show the amplitudes of the terms that are most likely to be detectable, namely LSD/RSD, Doppler and lensing. On the left of
figure \ref{fig:kernelsV} we show the amplitude of both LSDs \eqref{eq:ALSD} and redshift space distortions (RSD). We can see that the redshift dependence of both amplitudes is quite different. Although the redshift and luminosity density perturbations are linear in the peculiar velocity when written in terms of background quantities (a step needed to compute the number counts fluctuation), their effect on the volume perturbation has a different amplitude and redshift dependence. We see an inversion of the strength of the LSD/RSD; 
LSDs are stronger than RSDs after $z\approx1.6$ (which agrees with results shown in \cite{Libanore2021}).

Notably, the Doppler term on the right of figure \ref{fig:kernelsV} is different altogether: while in RS it contributes strongly near the observer and then tends to a constant value close to zero, in LDS it is roughly constant, tending to \vnd{$-5$ at $z=0$ and slowly tending to 0} at high redshifts (see eq. \eqref{eq:AD}). Note that we fixed the biases $b_e=s=0$ but this does not change the overall trend, only the limits at low and high-z. For this particular choice of evolution and magnification biases, the amplitudes in the different spaces have \vnd{both negative} signs. Additionally the LDS Doppler term does not have an $1/r$ dependence near the observer. This is an effect of expanding the observed luminosity distance in terms of the background redshift, which brings
an extra factor of $\delta z$.
In fact, this carries an extra velocity term which eliminates the dependency on $1/r$, and that will enter in the expression for the luminosity distance perturbation, resulting in a non-zero Doppler term at high redshifts.

In figure \ref{fig:kernelsL} we plot the lensing coefficient in eq. \eqref{eq:AL}. On the left, we fix the source at redshift $5$ and look at $A_L$ as a function of $r(z)$ which is the integrand in eq. \eqref{eq:finalAs}. 
This implies that the overall trend of the redshift space curve won't change depending on the source redshift. However, the amplitude will depend on the value of $s$, which we have set to zero. In the case of LDS, the curve's shape and the zero crossing point will depend on the actual values of $b_e$ and $s$. This means that \vnd{tracers in LDS will have more structure in the line-of-sight lensing kernel with transitions from magnification to de-magnification, and/or vice-versa}%further away
, while in RS there is only a magnification effect. In the right-hand panel, we evaluate the lensing integrand at the halfway distance between source and observer, $r=0.5\bar r$, and examine how it changes with $\bar r(z)$ (i.e the source/background distance). While in redshift space we have a constant value, in LDS it depends on the location of the source. 

\begin{figure}
    \centering
    \includegraphics[width=\textwidth]{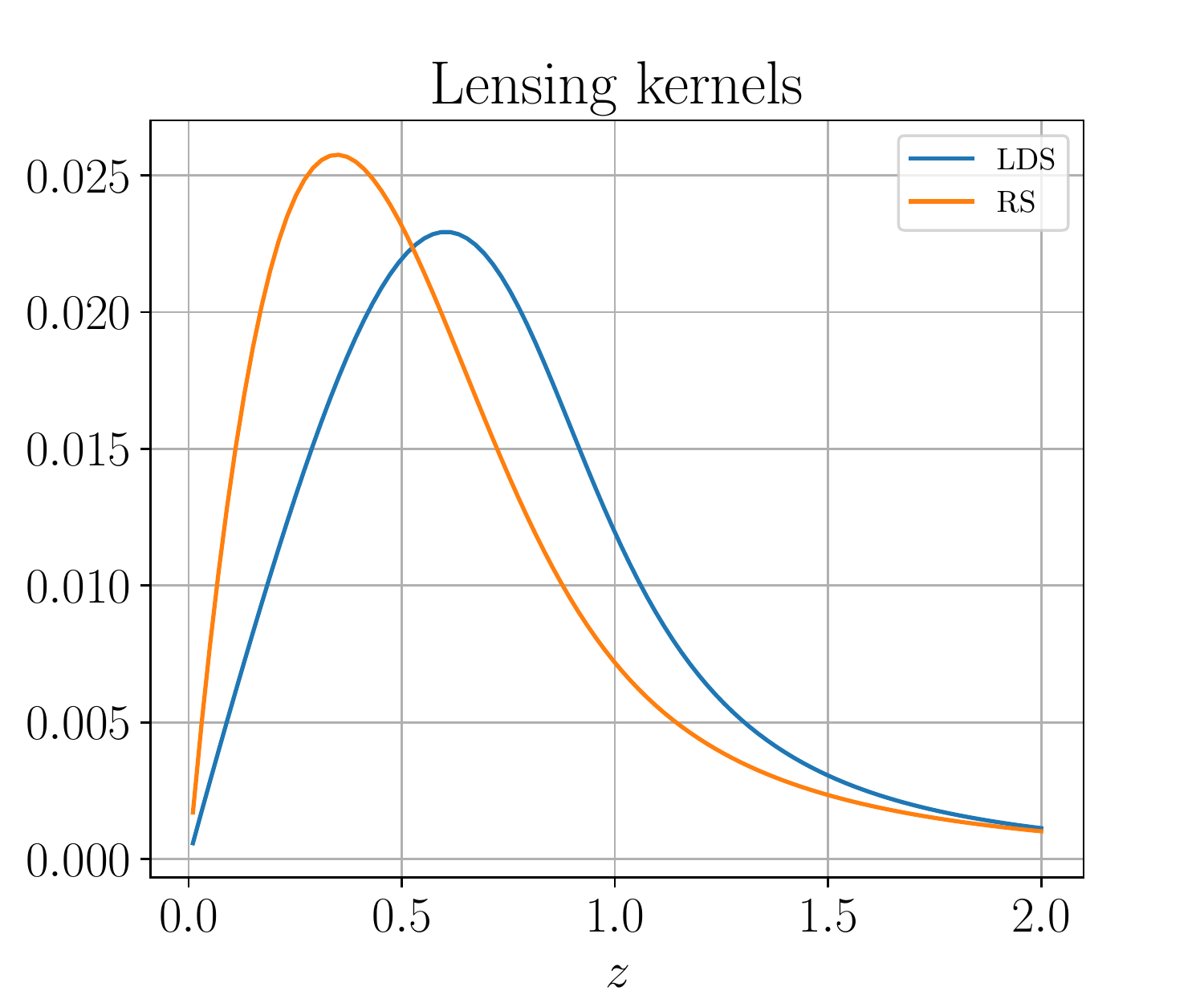}
    \caption{Comparison of the {normalised} lensing kernels from eq. \eqref{eq:lensing_kernel} in luminosity distance space (blue) and redshift space (orange).}
    \label{fig:lensing}
\end{figure}

This difference can be better seen when using the lensing kernel commonly used for weak lensing.
 Weak gravitational lensing is encoded in the convergence field $\kappa_c$, which can be expressed as a weighted projection of the matter overdensity \cite{Bartelmann_2001,Hand_2015}. In redshift space:
\bea
\kappa_c(\boldsymbol{\hat{n}}) = \int_0^{\infty} \d z\ W_L(z)\delta(r(z),\boldsymbol{\hat{n}},z) \, ,\
\eea
where, assuming a flat Universe, the lensing kernel $W_L$ depends on the amplitude $A_L$ from eq. \eqref{eq:AL} via:
\bea\label{eq:lensing_kernel}
W_L(z) = \frac{3}{2}\Omega_m \frac{H_0^2}{\mathcal{H}(z)}\frac{r(z)}{c}\int_z^{\infty} \d z_s\ A_L(r(z_s))\ p(z_s) \, ,\
\eea
with $p(z_s)$ being a normalised distribution of tracers.
In figure \ref{fig:lensing}, we plot a comparison of the lensing kernel in luminosity distance space and in redshift space. We adopt a fitting function from \cite{Hand_2015} parameterising the redshift distribution of the galaxy sample from the Canada-France-Hawaii Telescope Stripe 82 Survey \cite{Erben_2013,Miller_2013}:
\bea
    p(z_s)=A\frac{z^a+z^{ab}}{z^b+c} \, ,\
\eea
with $A=0.688$, $a=0.531$, $b=7.810$ and $c=0.517$. This is simply a toy model to compare the lensing kernel in luminosity distance to previous work in redshift space. {In LDS the kernel is broader and thus lensing is sensitive to a wider range of redshifts. The particular choice of distribution of sources will also affect this, together with adding the impact of magnification and evolution biases. In particular, for a specific choice of distribution the resulting kernel can change sign.}
%It is clear that in LDS one is more sensitive to the structure at lower redshifts, mainly due to the correction term present in eq. \eqref{eq:AL}, causing a sharper peak at low $z$. The dip present in LDS likely arises from the zero crossing previously shown in figure \ref{fig:kernelsL} and the particular choice of distribution of sources. In particular, for a specific choice of distribution the resulting kernel can change sign.

\section{The angular power spectrum in luminosity distance space} 
\label{sec:gr_effects}
\subsection{A comparison with power spectra in redshift space}

In this section we use the full expression shown in eq. \eqref{eq:num_final} to evaluate numerically the angular power spectrum, and the relevance of the relativistic corrections. We modified the publicly available code \texttt{CAMB}\footnote{https://github.com/cmbant/CAMB} together with {its python wrapper \texttt{pyCAMB}, and} implemented the option to calculate the auto and cross-correlation angular power spectra in luminosity distance space, {including correlations between distinct bins in distance,} using the expression in \eqref{eq:general_num_contrast}.\footnote{We will make the code publicly available shortly.} In order to compute the angular power spectrum {of sources within a luminosity distance range $D_L \in [D_{L,i}-\Delta D_{L,i}/2,D_{L,i}+\Delta D_{L,i}/2]$, where $\Delta D_{L,i}$ is the size of the \emph{ith} bin,} we expand the number density contrast in luminosity distance space into spherical harmonics, i.e.,
\be
\De(\bn,D_{L,i})= \sum^{\infty}_{\ell = 0} \sum^{\ell}_{m = -\ell} \alm(D_{{L,i}}) \ylm{\bn}\,.
\ee
The angular power spectrum $C_\ell$ is the two-point function, equivalent to the variance of the $\alm$ coefficients, and is related with the primordial power spectrum ${\cal P}_{\cal R}$ via
\be
C^{ij}_\ell=\langle\alm(D_{L,i}) a^{\ast}_{\ell'm'}(D_{L,j}) \rangle= 4\pi \int {\rm d}\ln k\ \De^i_{\ell}(k,D_{L,i})\De^j_{\ell}(k,D_{L,j})\ {\cal P}_{\cal R}(k)~ \delta_{\ell\ell'}\delta_{mm'} \, ,
\ee
where $\De^i_\ell$ is the transfer function of the \emph{ith} bin and combines the distribution of sources, the window function and the transfer functions of each term in eq. \eqref{eq:finalAs}. {Here $i$ and $j$ stand for sample bins in luminosity distance and/or redshift, which can be distinct.} A comprehensive derivation of the transfer functions in the bin in harmonic space is found in Appendix \ref{sec:LDharm}. %To summarise, 

\begin{figure}
    \centering
    \includegraphics[width=0.8\textwidth]{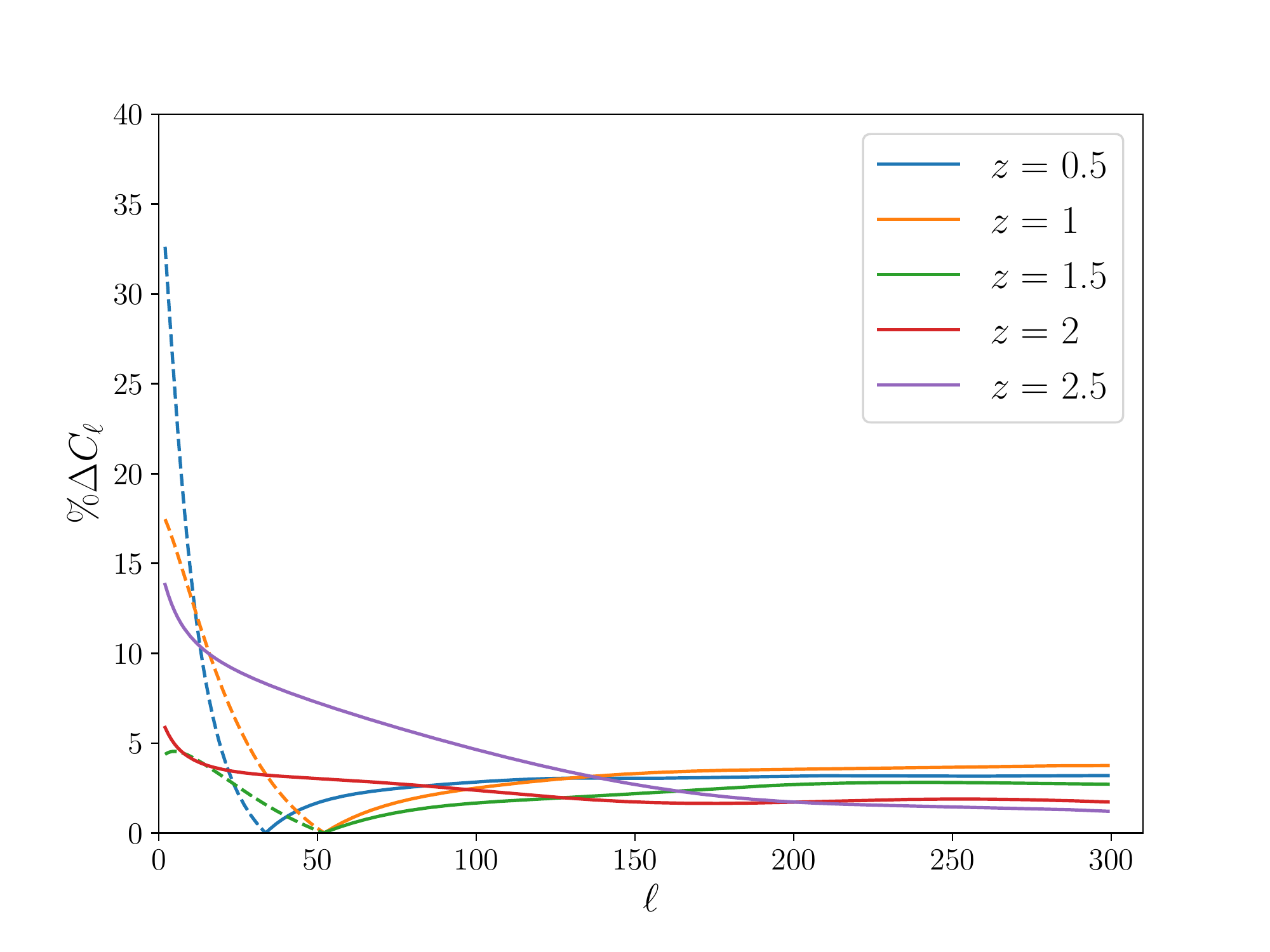}
    \includegraphics[width=0.8\textwidth]{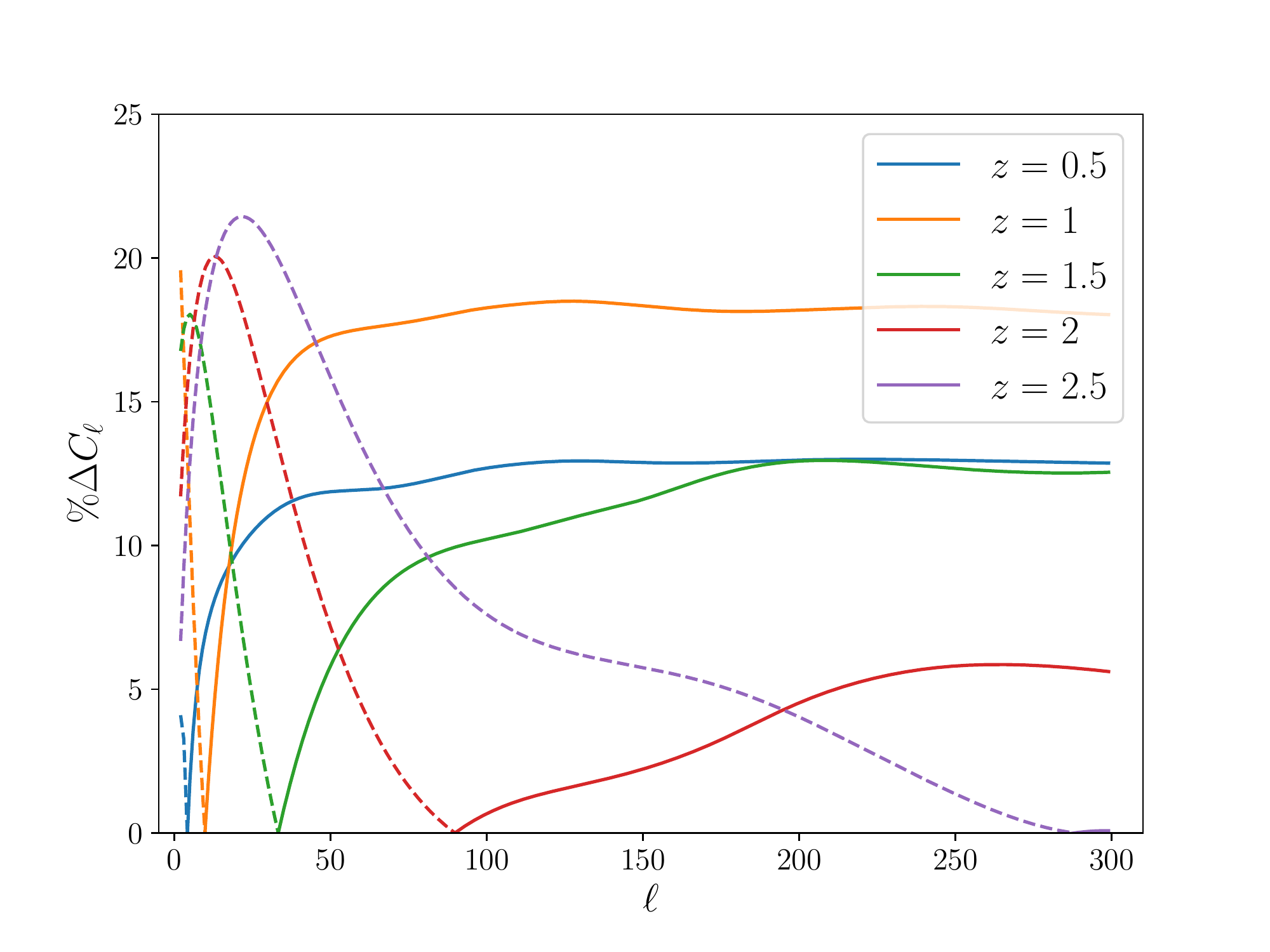}
    \caption{Percentage difference between the angular power spectrum in luminosity distance space (LDS) and redshift space (RS) for two different binning sizes (\textit{top}: $\sigma=0.05$, \textit{bottom}: $\sigma=0.2$). At low $\ell$ the two can differ quite significantly, particularly at high redshift. However, at smaller scales the difference tends to a constant depending on the redshift examined. The overall difference is also related to the binning used, thus on the uncertainty on the localisation of the tracer, as better accuracy on the distance will reduce the difference between the angular power spectra in the two spaces.}
    \label{fig:dl_to_z_diff}
\end{figure}

We start from the expression of the overdensity (eq. \eqref{eq:finalAs}) and the definition of $C_\ell$ to construct $\De^i_\ell$, as well as the theoretical harmonic space transfer functions for each contribution in eq. \eqref{eq:finalAs}. These are then implemented in \texttt{CAMB}.
The code also allows the input of the values of both magnification and evolution bias, although for the following examples we choose to set them to zero. {Note that \cite{Libanore2022} also changed \texttt{CAMB} to provide a code to compute the angular power spectra in luminosity distance space. We go beyond their work and include all contributions present in eq. \eqref{eq:finalAs} without the Limber approximation. We also include the magnification bias in our calculation and code. The latter also allows for cross-correlations between luminosity distance and redshift spaces.}

Before looking at the individual contributions, we compare the angular power spectrum in the two different spaces. {For simplicity we will consider  \textit{generic} tracers throughout this paper that live in either space with idealised values for its biases, i.e., $b=1, s=b_e=0$. Therefore, one can directly compare the expressions in luminosity distance space (LDS) and redshift space (RS).} In figure \ref{fig:dl_to_z_diff} we plot the relative difference between the angular power spectrum {in LDS and the one in RS} at multiple redshifts. We use
broad Gaussian windows in the top plot and narrow bins in the bottom plot. Whilst broad windows are motivated by the large uncertainty in the estimation of the luminosity distance for GWs \cite{GWTC2,GWTC3}, narrower bins are used in galaxy clustering and thus are appropriate for a comprehensive comparison. Further, SNIa distance uncertainties for LSST will lie in between these two regimes \cite{Pantheon,Howlett_2017}.

Interestingly, using large bins (bottom figure \ref{fig:dl_to_z_diff}), the large scales of the angular power spectrum can experience a \vnd{$20\%$} difference at $z=2.5$. But at smaller scales all angular power spectra tend to a constant which depends on redshift. At low redshift, the difference is roughly $\sim 10\%$, implying a relatively small difference between the two cases when looking at smaller scales. These differences are reduced significantly when looking at thinner bins (top figure \ref{fig:dl_to_z_diff}), however, we remind the reader that the more realistic scenario when using GWs/SNIa would need larger uncertainties, and thus would be more akin to the bottom plot. As we will see in figures \ref{fig:comparison_dl_space} and \ref{fig:comparison_z_space}, the bulk of the difference at low $\ell$ is due to the lensing contribution.
Instead, at higher $\ell$ the two main contributions are the density auto-correlation (which is identical in the two spaces), and LSD/RSD terms; in particular, the latter gives a constant difference between LDS and RS, which is then visible at smaller scales when the differences in all other contributions become negligible.

We note that this constant offset at high $\ell$ \vnd{increases towards $z \sim 1.0$ and decreases} afterwards. Looking back at figure \ref{fig:kernelsV}, the amplitude of the LSDs becomes stronger than that of the RSDs \vnd{after} this same value of redshift. In $\Lambda$CDM, $z_t\simeq1.5$ is the turnover point for the angular diameter distance, and assuming no shear in a preferential direction, to a safe approximation, the angular diameter distance is the same as the area distance \cite{Fleury_2015} used to identify the perturbation in luminosity distance (in eq. \eqref{eq:lumdist_start}). The imprint of the turnover is then visible in how the difference in the angular power spectra between LDS and RS evolves with redshift. However, this behaviour only appears when considering displacement-related effects, i.e. LSD/RSD, Doppler and ISW. We suggest this is the effect of the extra $\delta z$ term in the last line of eq. \eqref{eq:lumdist_start}, as discussed at the end of section \ref{sec:expression}.  

%Duplicated
%This constant offset at large $\ell$ decreases towards $z \sim 1.5$ and increases afterwards. Looking back at figure \ref{fig:kernelsV}, the amplitude of the LSDs becomes stronger than that of the RSDs around this same value of redshift. In $\Lambda$CDM, $z_t\simeq1.5$ is the turnover point for the angular diameter distance, and assuming no shear in a preferential direction, to a safe approximation, the angular diameter distance is the same as the area distance \cite{Fleury_2015} used to identify the perturbation in luminosity distance (in eq. \eqref{eq:lumdist_start}). The imprint of the turnover is then visible in how the difference in the angular power spectra between LDS and RS evolves with redshift. However, this behaviour only appears when considering displacement-related effects, i.e. LSD/RSD, Doppler and ISW. We suggest this is the effect of the extra $\delta z$ term in the last line of eq. \eqref{eq:lumdist_start}, as discussed in section \ref{sec:expression}.  

\begin{figure}
    \centering
    \includegraphics[width=0.48\textwidth]{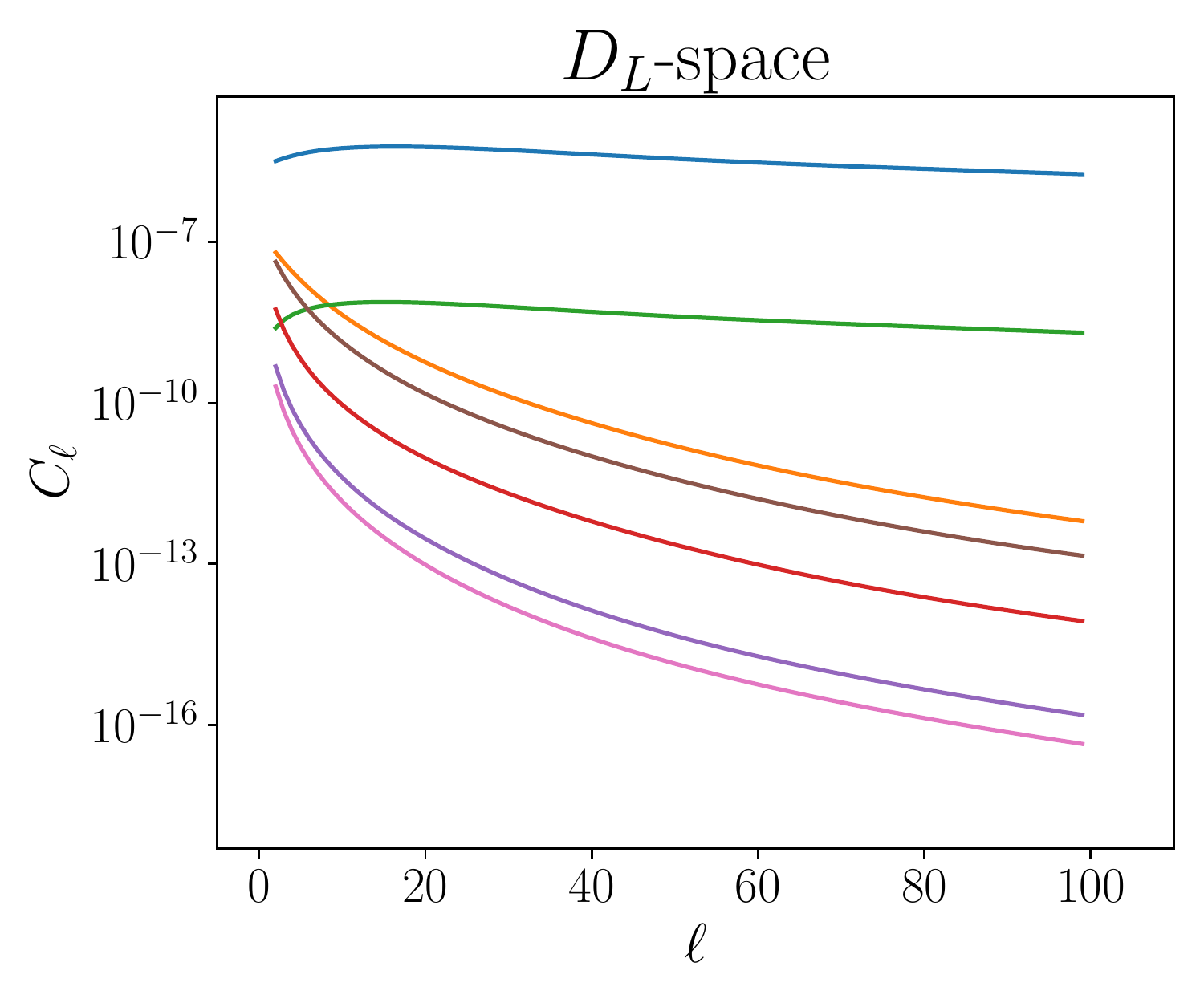}
    \includegraphics[width=0.48\textwidth]{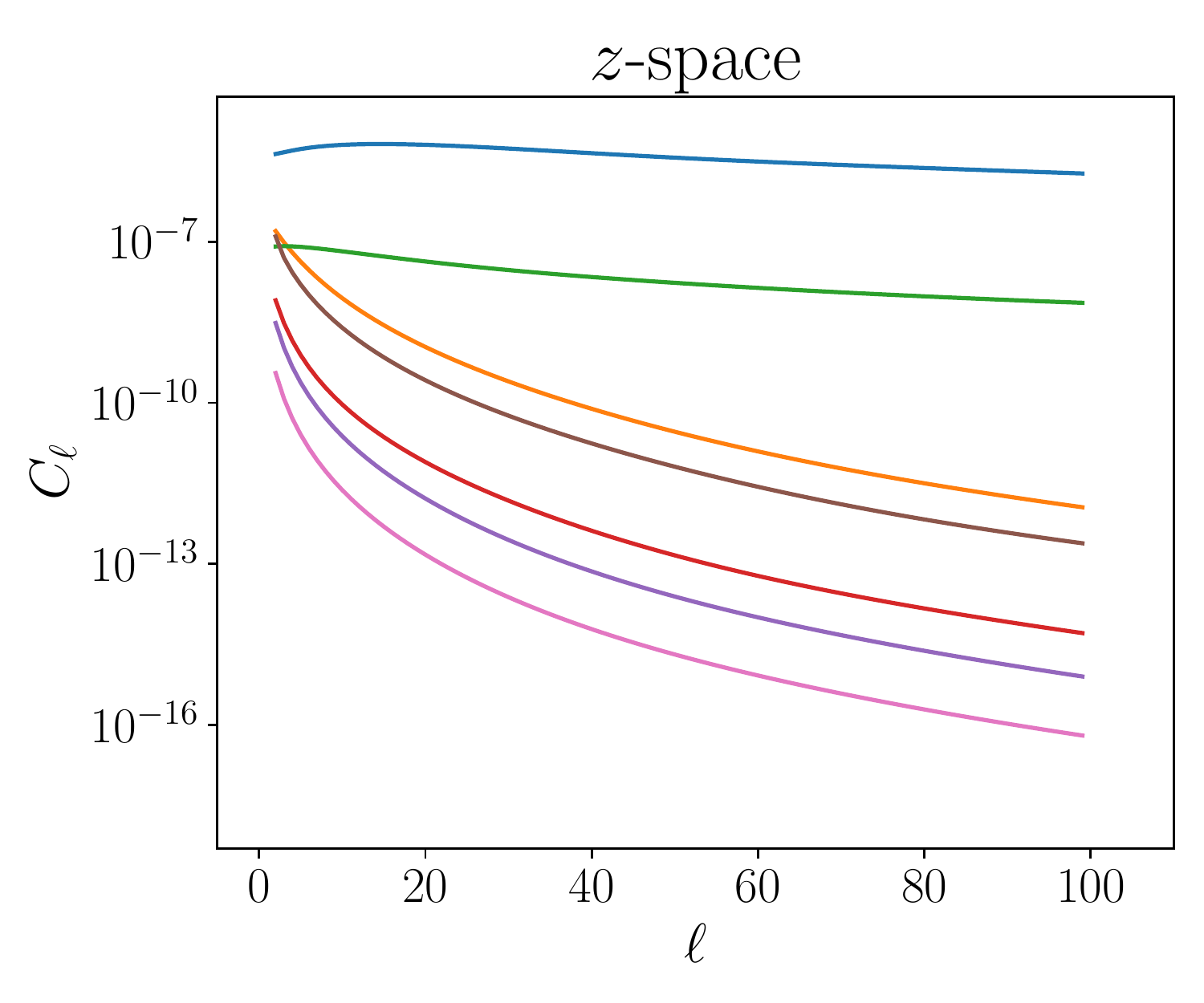}
    \includegraphics[width=0.48\textwidth]{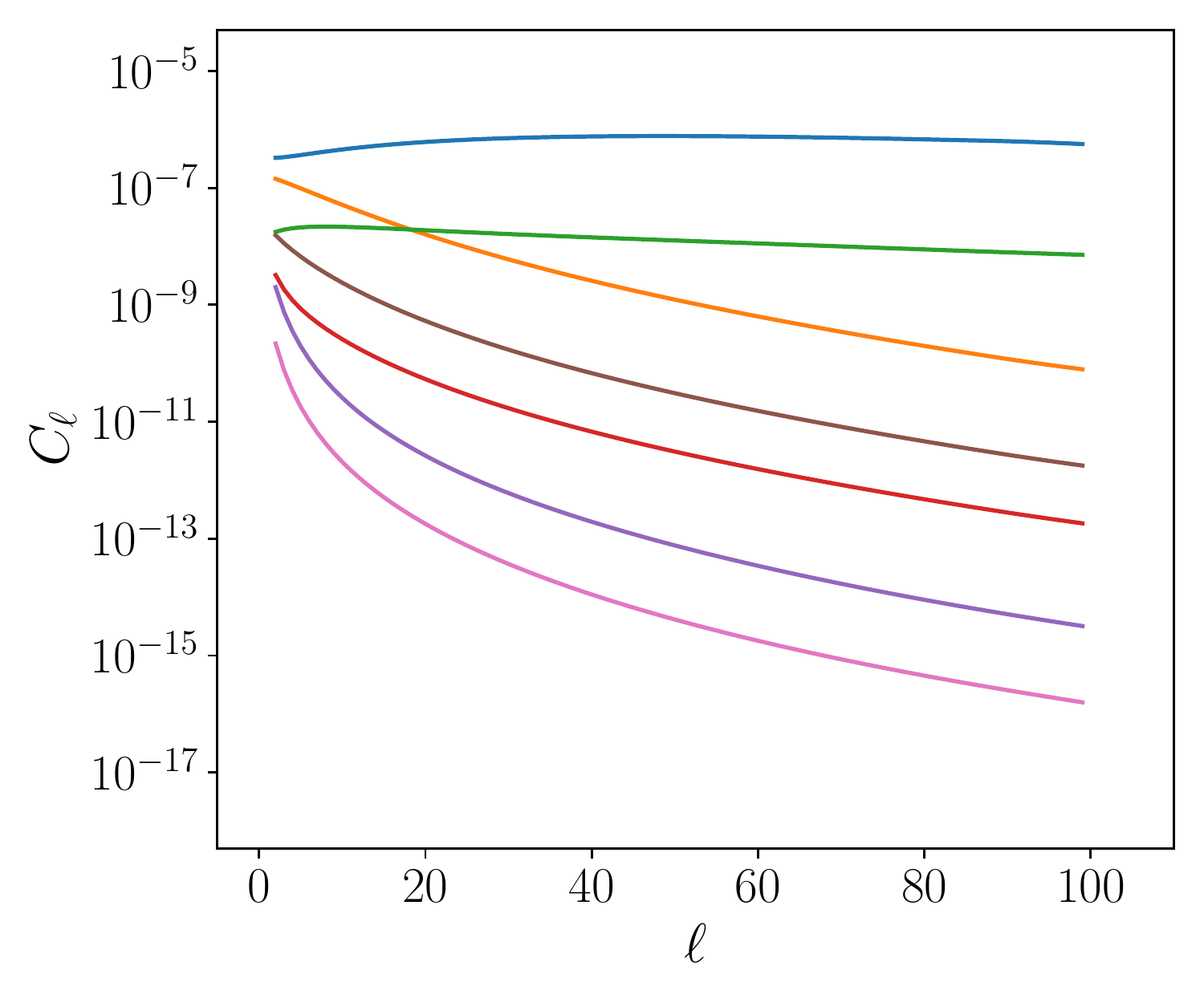}
    \includegraphics[width=0.48\textwidth]{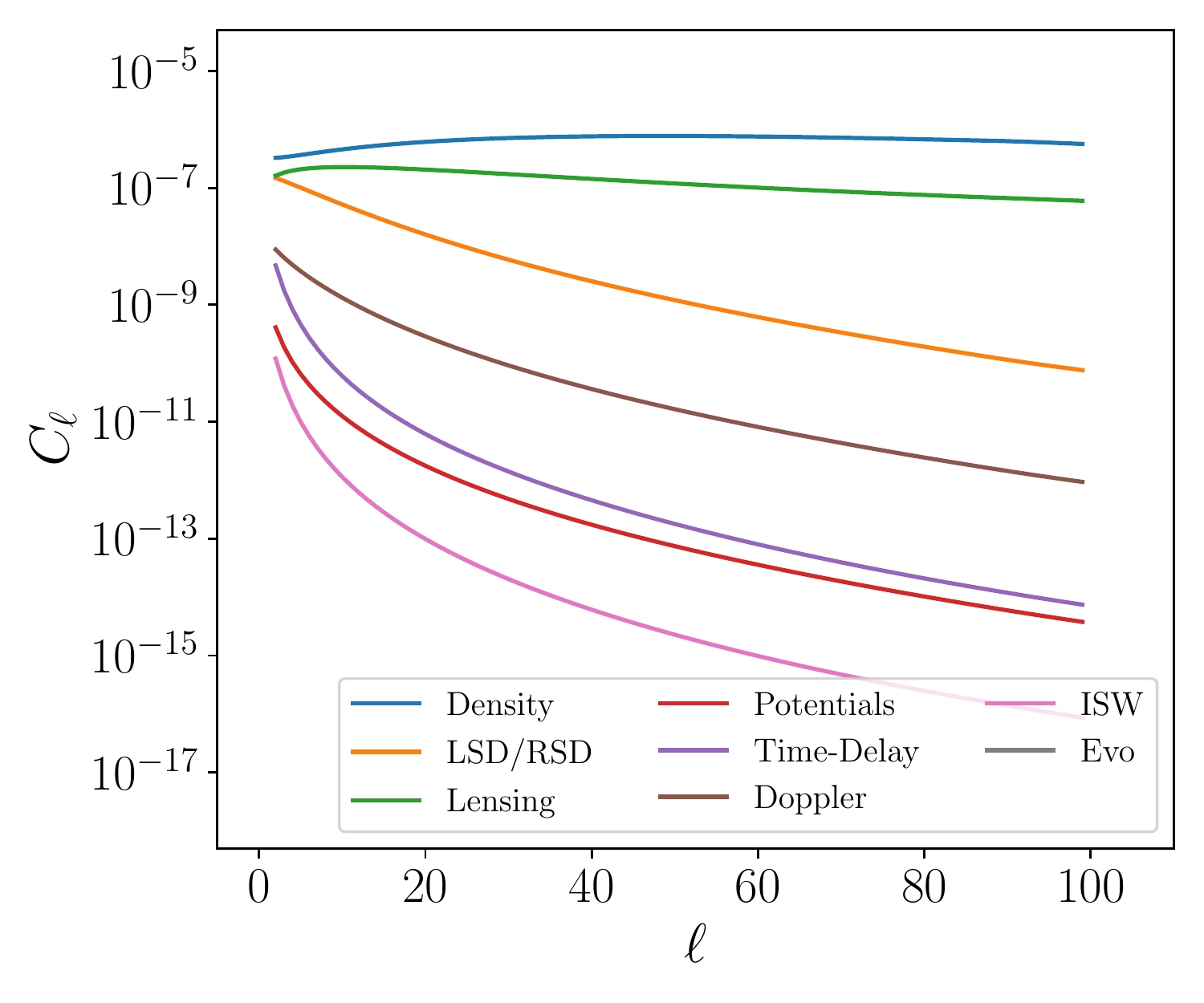}
    \caption{Angular power spectrum {for the generic tracers (see text)} at $z=0.5$ (\textit{top}) and at $z=1.5$ (\textit{bottom}) and the auto-correlations of different contributions. {\em Left:} in luminosity distance space. {\em Right:} in redshift space.}
    \label{fig:comparison_1}
\end{figure}

More fundamentally, what figure \ref{fig:dl_to_z_diff} shows is that the number counts expression in redshift space is not a good approximation to estimate clustering of GW mergers, which are intrinsically observed in LDS.
One should note that the poor angular resolution of such experiments imposes in effect a beam, which when multiplied by the $C_\ell$ dampens the signal at high $\ell$. Here we did not model this effect in detail, as it is beyond the scope of this paper.

\subsection{The relevance of relativistic corrections}

We have seen that the angular power spectrum has a different amplitude in the two spaces, but we have not identified the origin of such a difference. In figure \ref{fig:comparison_1} we compare the angular power spectrum in luminosity distance space, including the auto-correlations of each contribution in the density contrast in equation \eqref{eq:num_final}, with the one in redshift case. We plot the $C_{\ell}$ up to $\ell_{max}=100$ considering the error in localisation of a GW event for 3G detectors like ET or CE \cite{Sathyaprakash_2011,  Sathyaprakash_2012, Scelfo_2018, Scelfo_2020, Scelfo_2022, Scelfo_2022_2} and that $\ell_{max}=180^{\circ}/\theta$, where $\theta$ is the angular uncertainty \cite{Mukherjee_2020}. At first glance, \vnd{one could say that lensing has a weaker} contribution in LDS as opposed to RS. %, together with other terms such as the potentials. 
However the other contributions are not so easily compared, especially because of the log scale of the plot. Nonetheless, we can identify the most relevant terms: lensing (green line) and velocity gradient (orange line).

\begin{figure}
    \centering
    \includegraphics[width=0.48\textwidth]{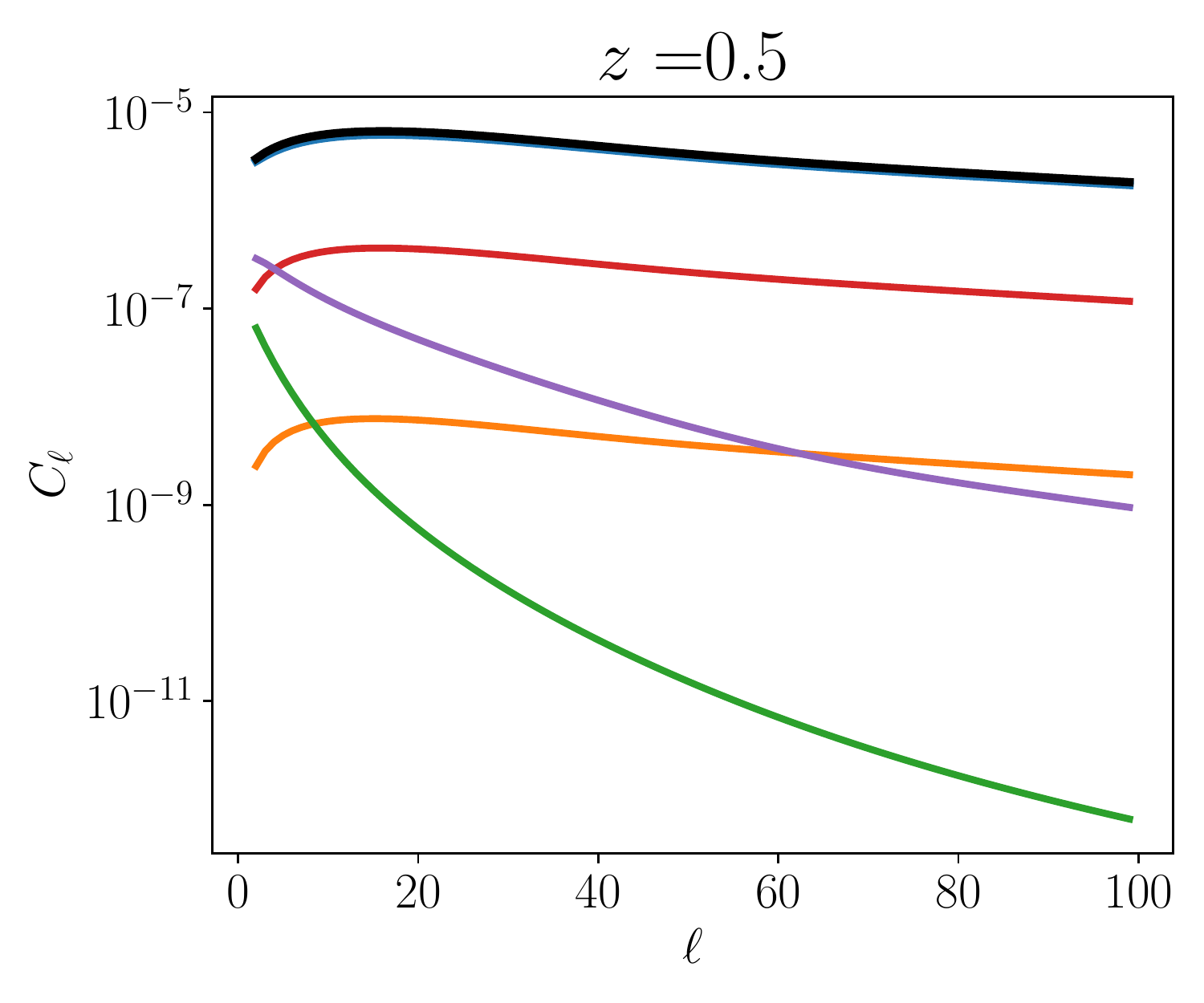}
    \includegraphics[width=0.48\textwidth]{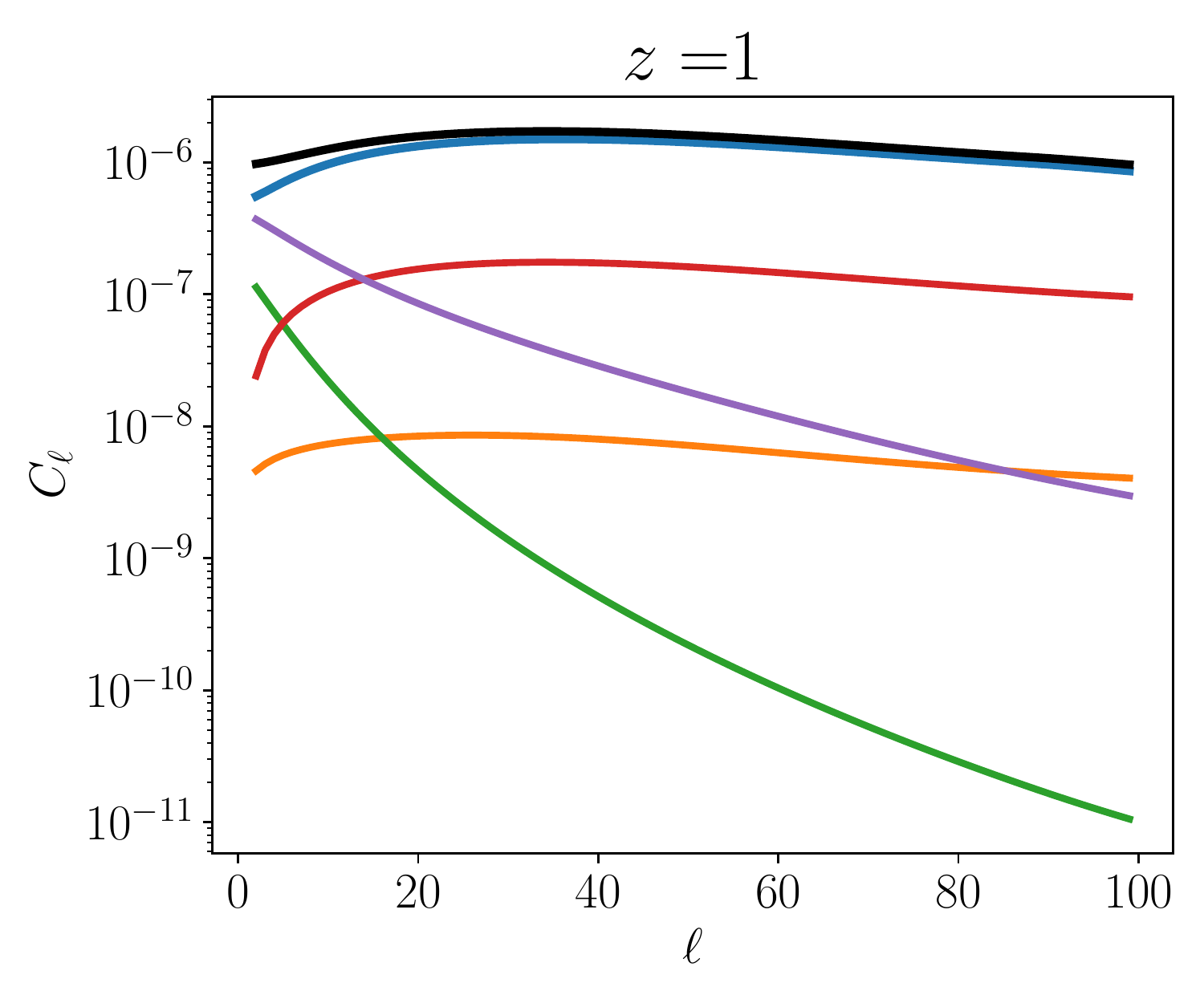}
    \includegraphics[width=0.48\textwidth]{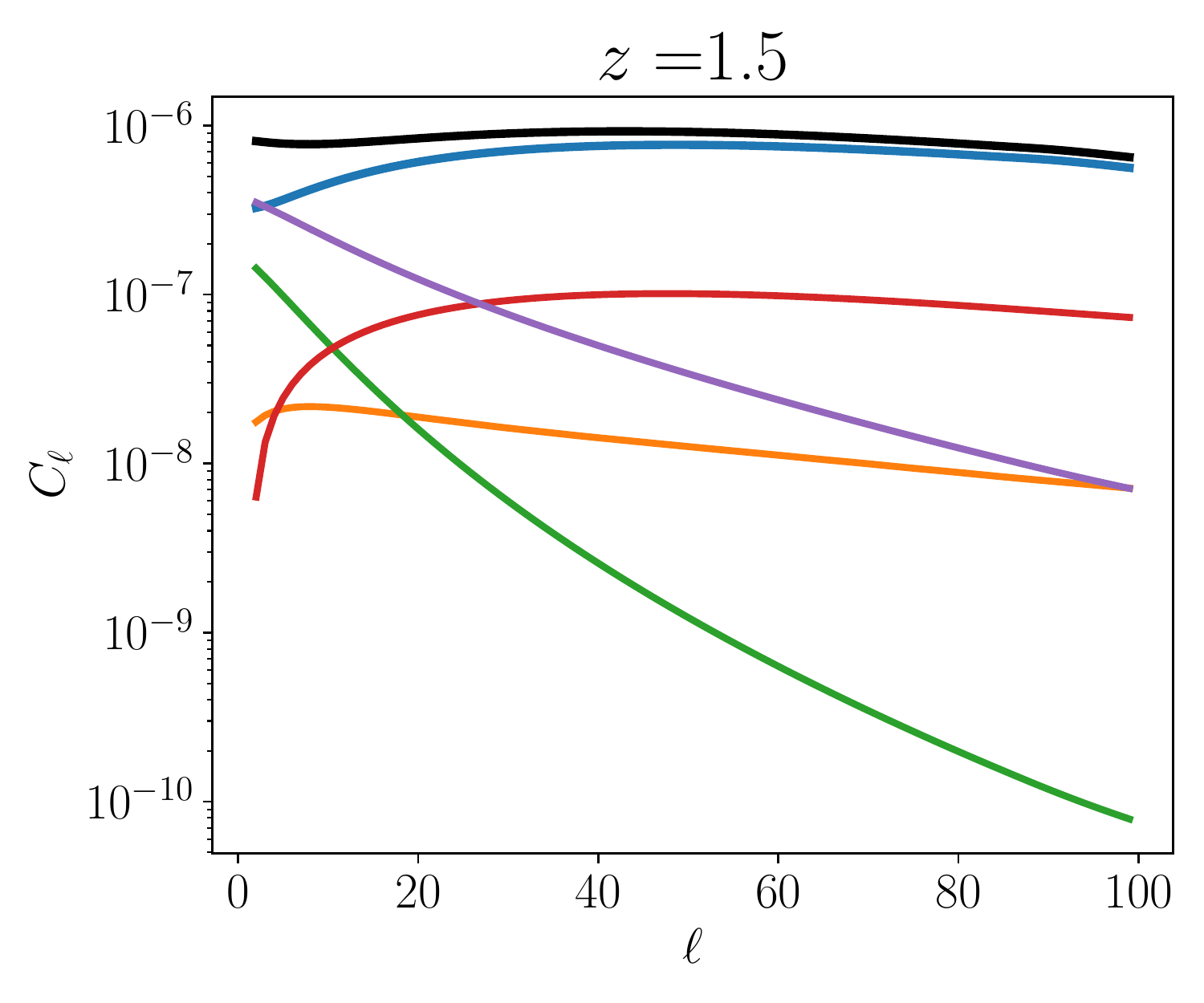}
    \includegraphics[width=0.48\textwidth]{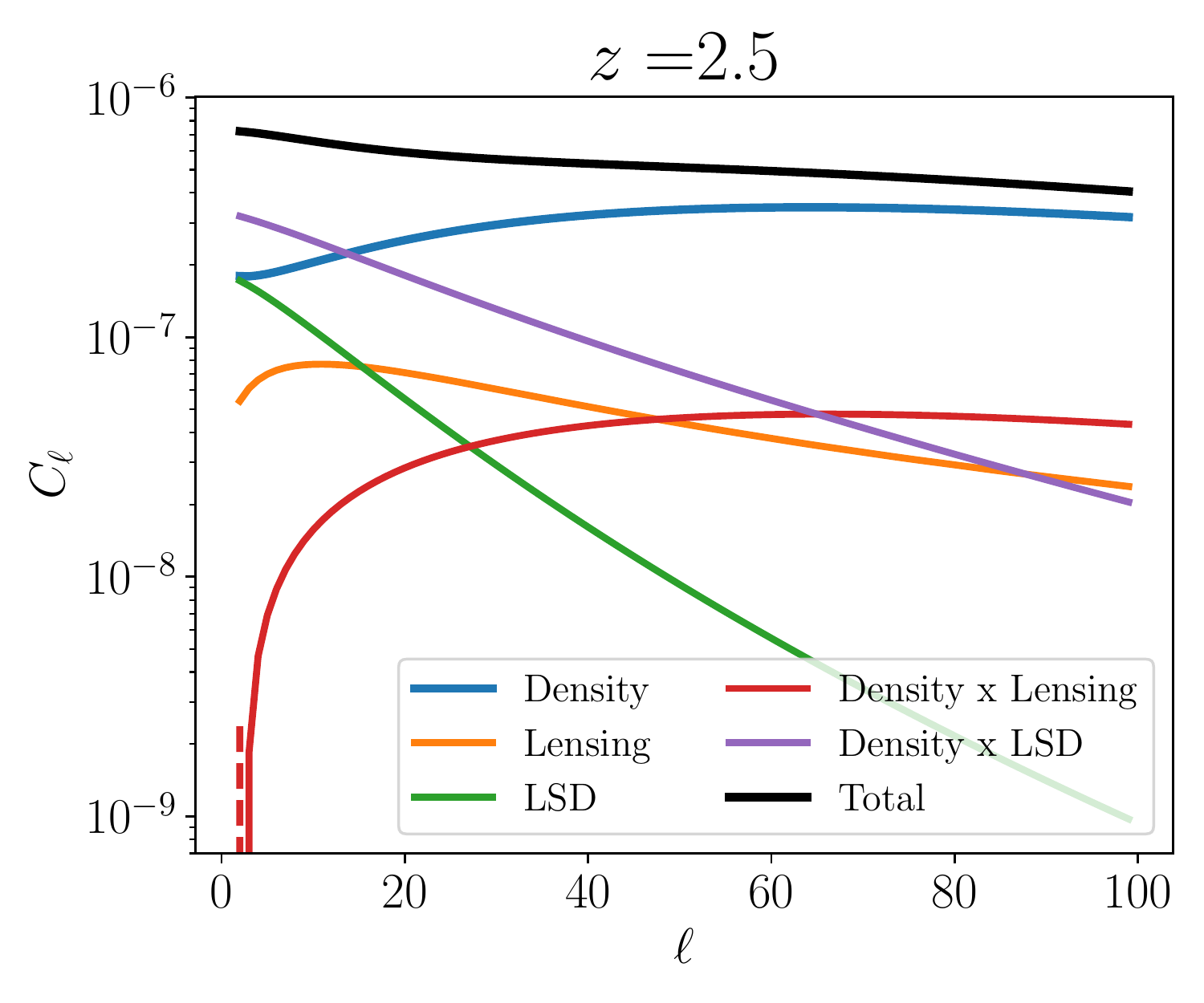}
    \caption{Comparison of the auto- and cross-terms contributions to the total angular power spectrum in luminosity distance space {for a generic tracer (see text)} at different redshifts. The higher the redshift and the more the lensing term contributes to the total.}
    \label{fig:comparison_dl_space}
\end{figure}

\begin{figure}
    \centering
    \includegraphics[width=0.48\textwidth]{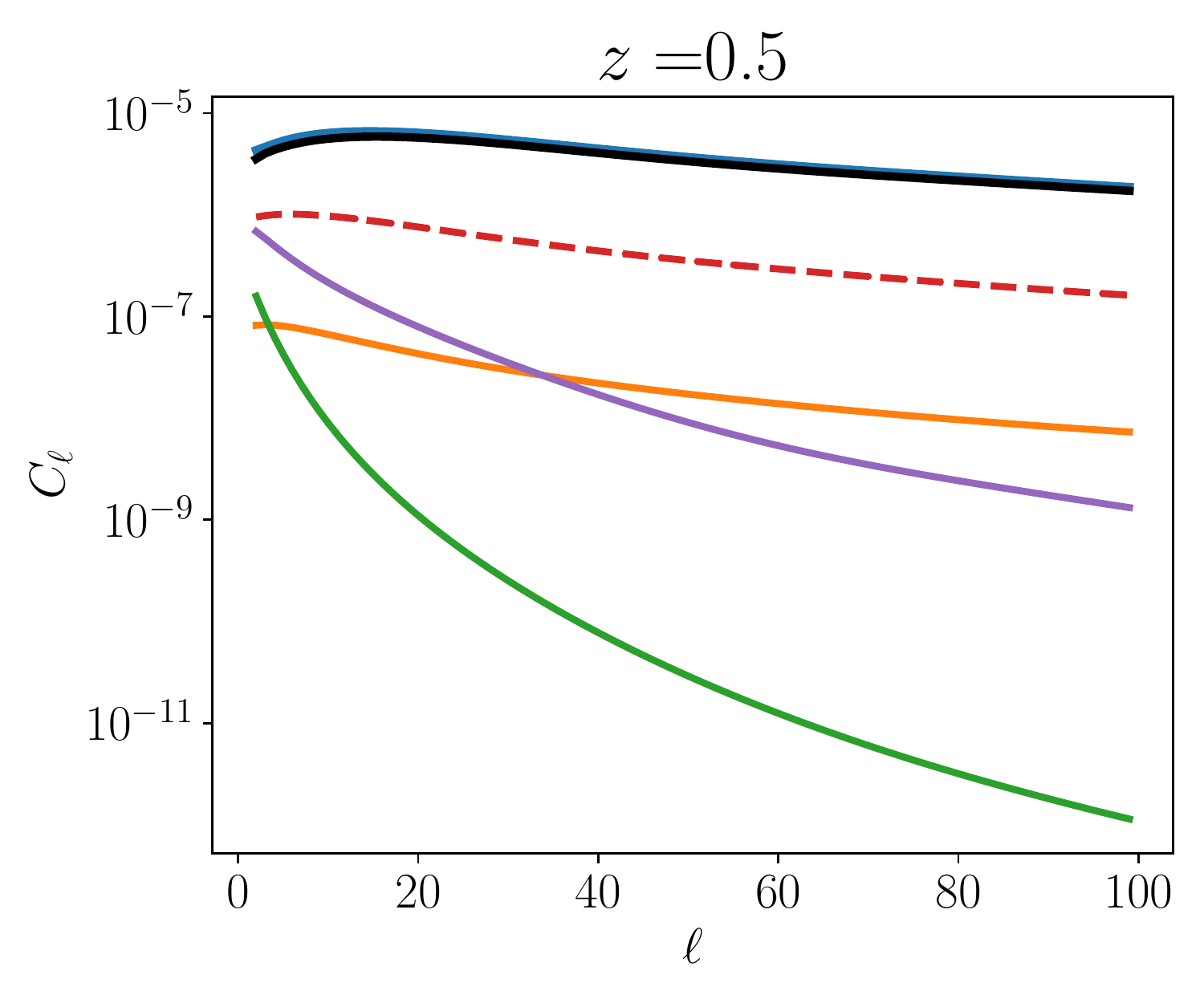}
    \includegraphics[width=0.48\textwidth]{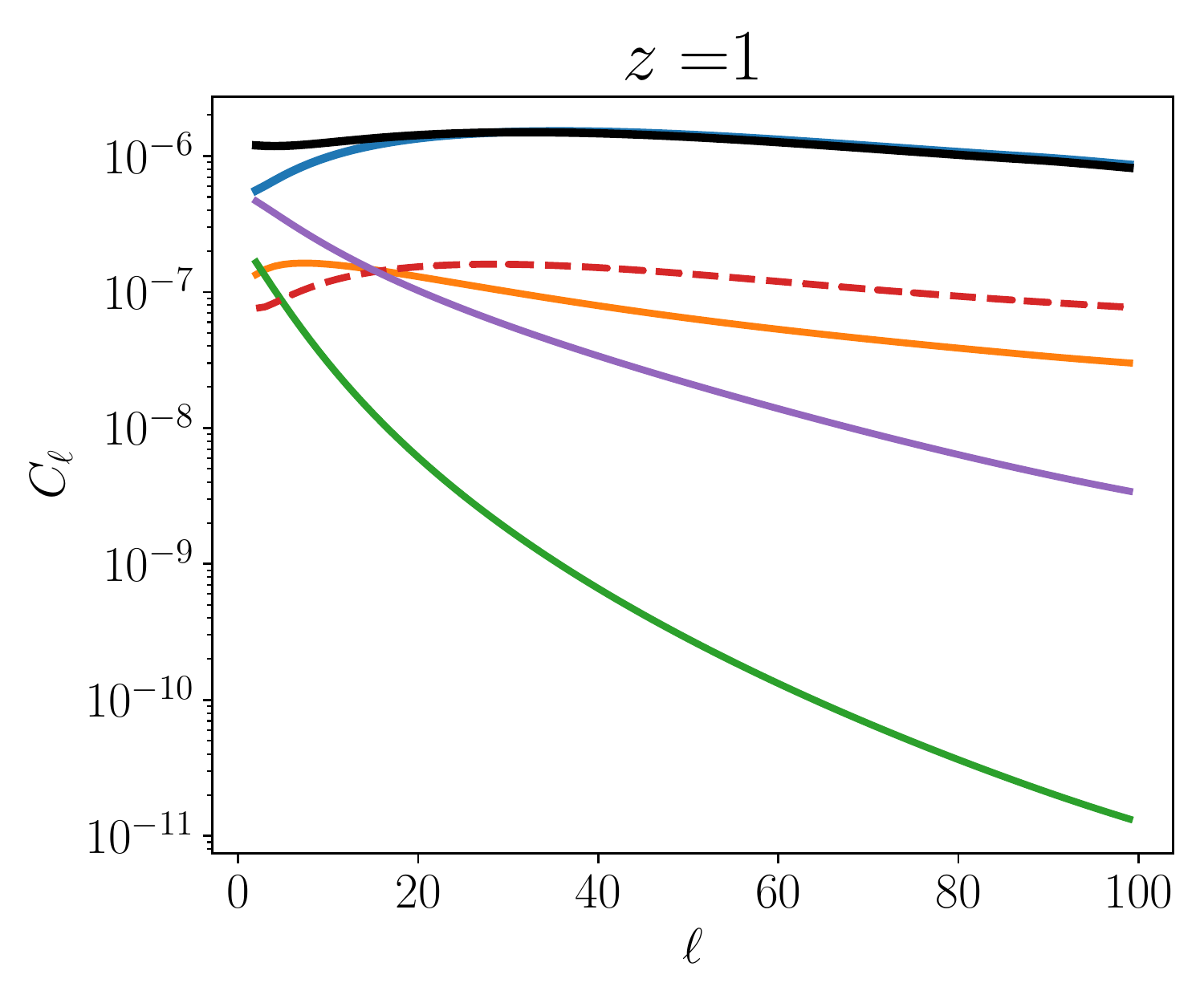}
    \includegraphics[width=0.48\textwidth]{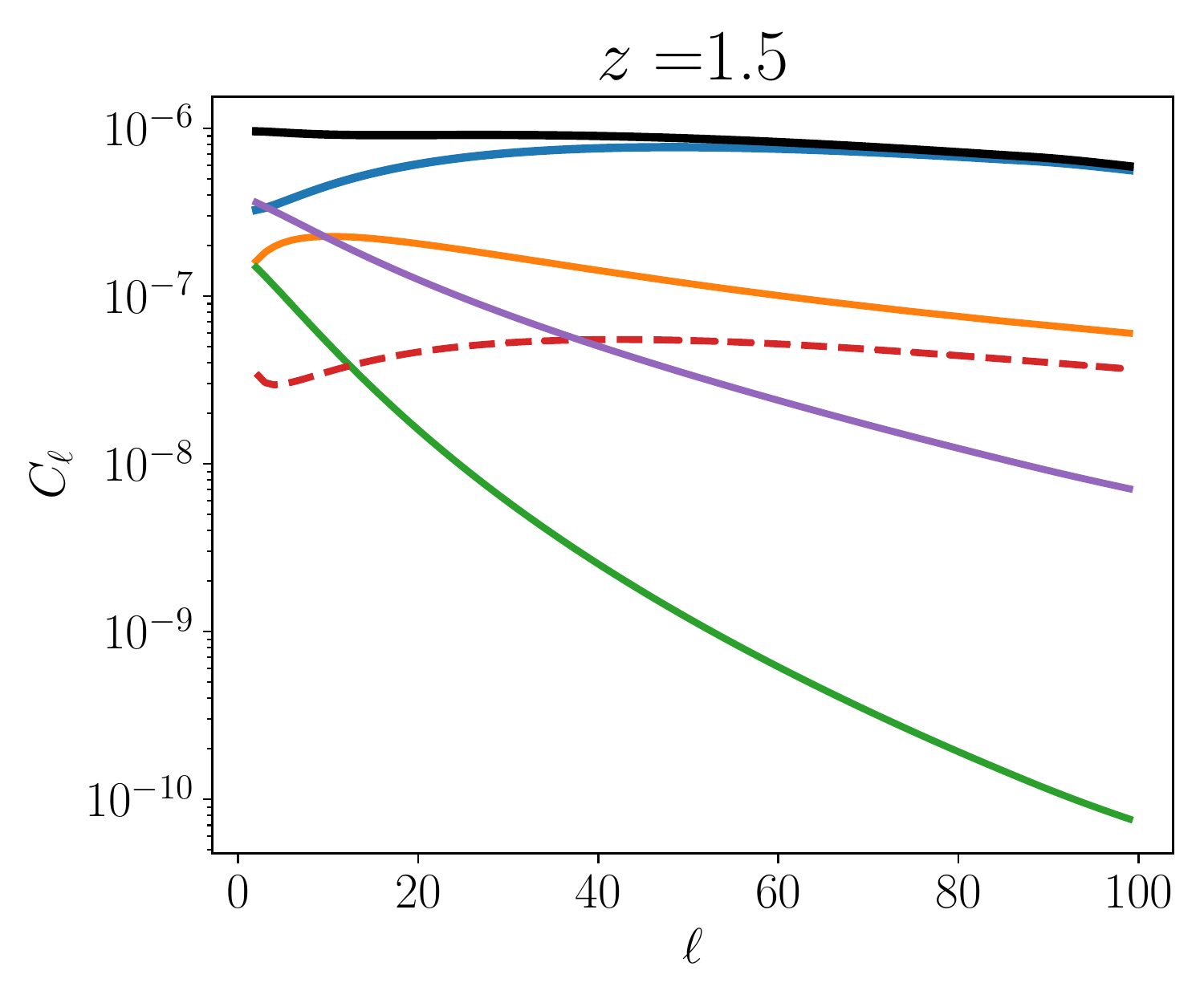}
    \includegraphics[width=0.48\textwidth]{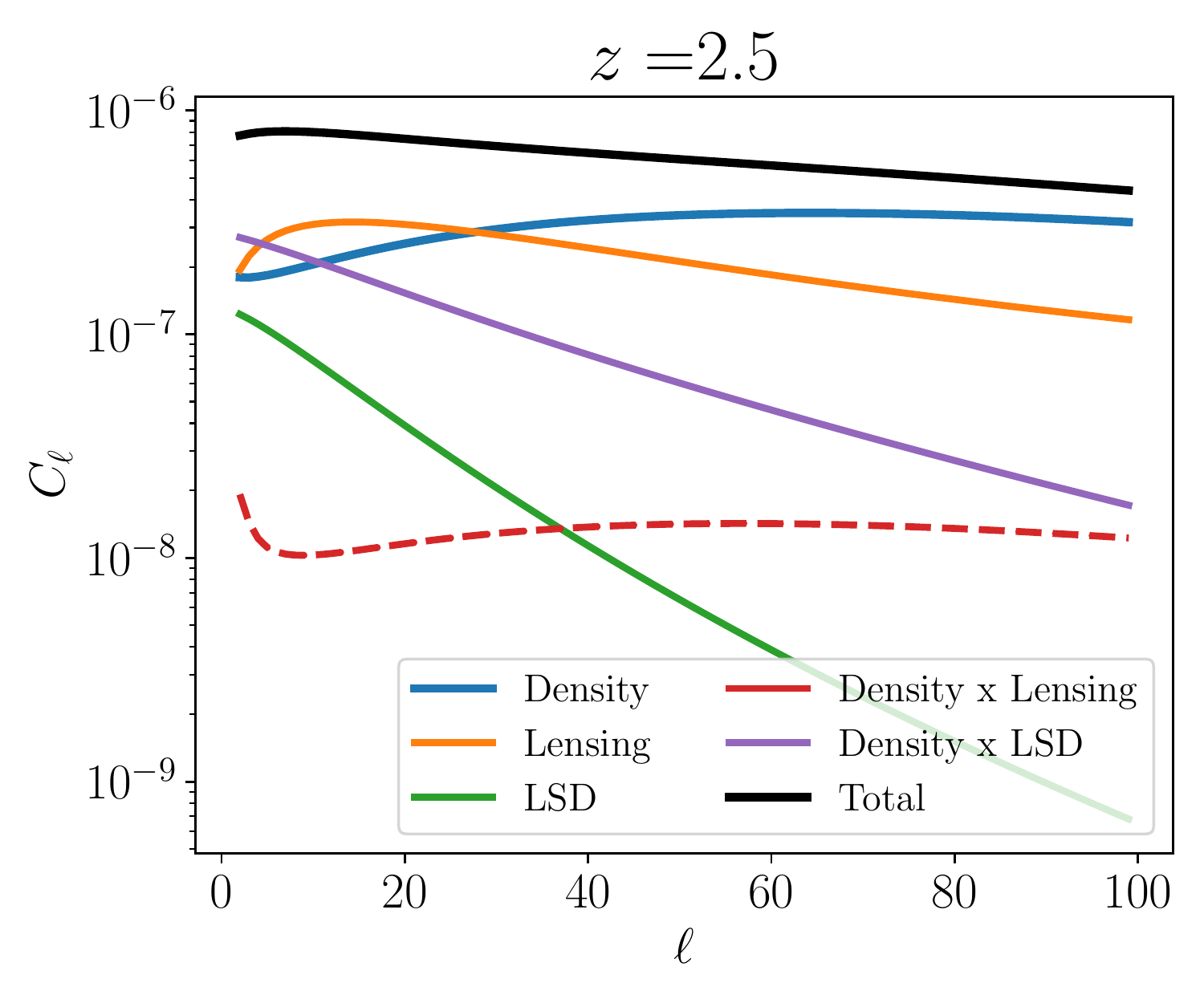}
    \caption{Comparison of the auto- and cross-terms contributions to the total angular power spectrum in redshift space {for a generic tracer (see text)} at different redshifts. Looking at figure \ref{fig:comparison_dl_space} we note that the lensing term in z-space is weaker at higher redshift compared to the LDS case.}
    \label{fig:comparison_z_space}
\end{figure}

To better gauge the relative importance of the main terms, we then plot in figure \ref{fig:comparison_dl_space} their contributions to the total angular power spectrum in $D_L$-space including the main auto- and cross-term correlations at several redshifts. As expected, the lensing auto-correlation contribution becomes more important as redshift increases. However, comparing with redshift space (figure \ref{fig:comparison_z_space}) we see that at high redshifts lensing has a stronger contribution in LDS as opposed to RS. In particular, we note that both lensing x lensing and density x lensing in LDS are greater than the RS case by roughly a factor of 2 when looking at $z=2.5$. This clearly shows this contribution to be the main difference between the two spaces, which then drives the percentage difference show in figure \ref{fig:dl_to_z_diff}. %In fact, the factor of 2 between the main contribution, i.e. lensing, is then reflected in a $50\%$ difference at low $\ell$ between the total angular power spectra in the two cases. 
Additionally the correlation between the density and lensing is always negative in the RS case, while in LDS it \vnd{mainly positive}.%can change sign due to the extra term in $A_L$.

\begin{figure}
    \centering
    \includegraphics[width=0.48\textwidth]{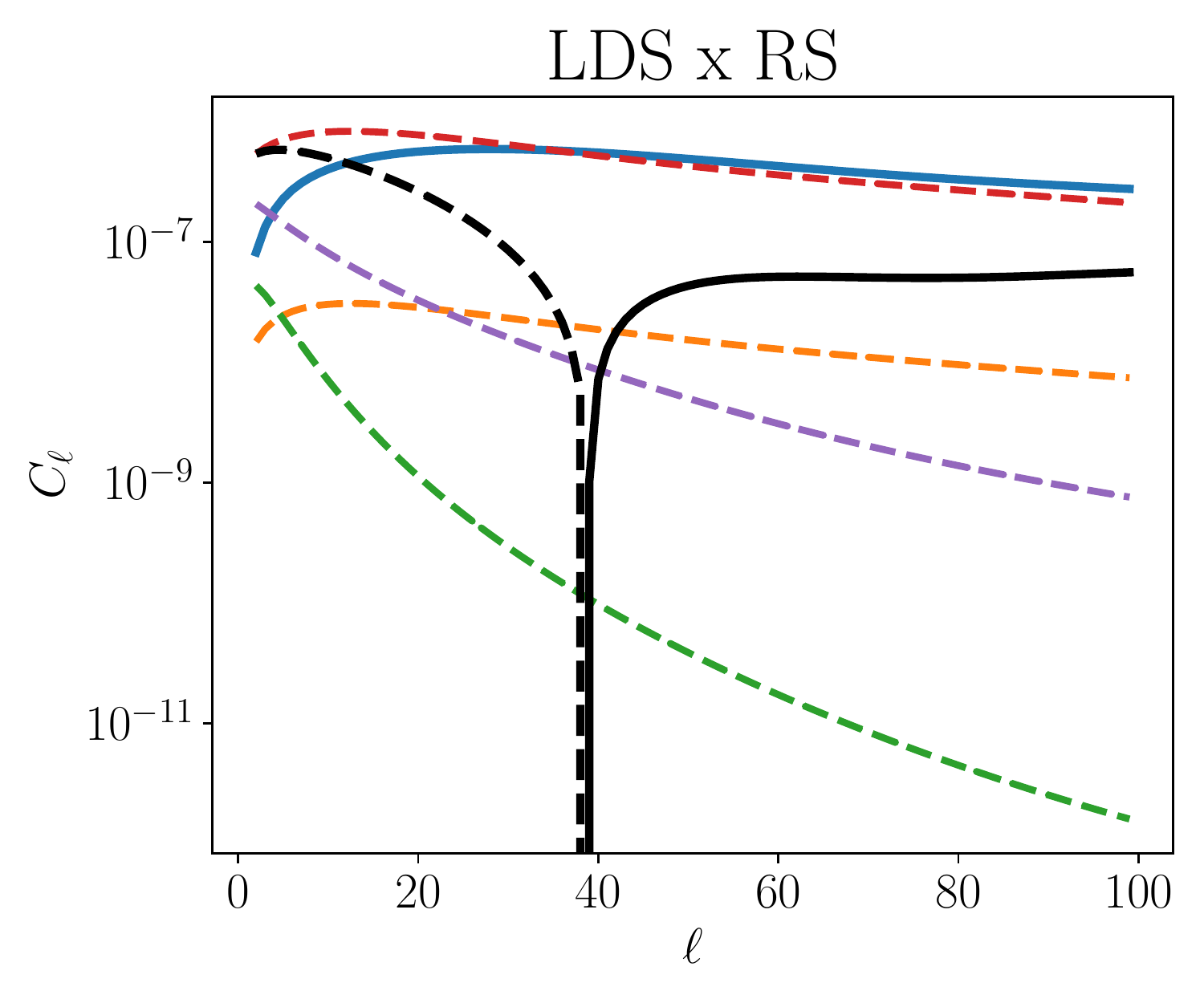}
    \includegraphics[width=0.48\textwidth]{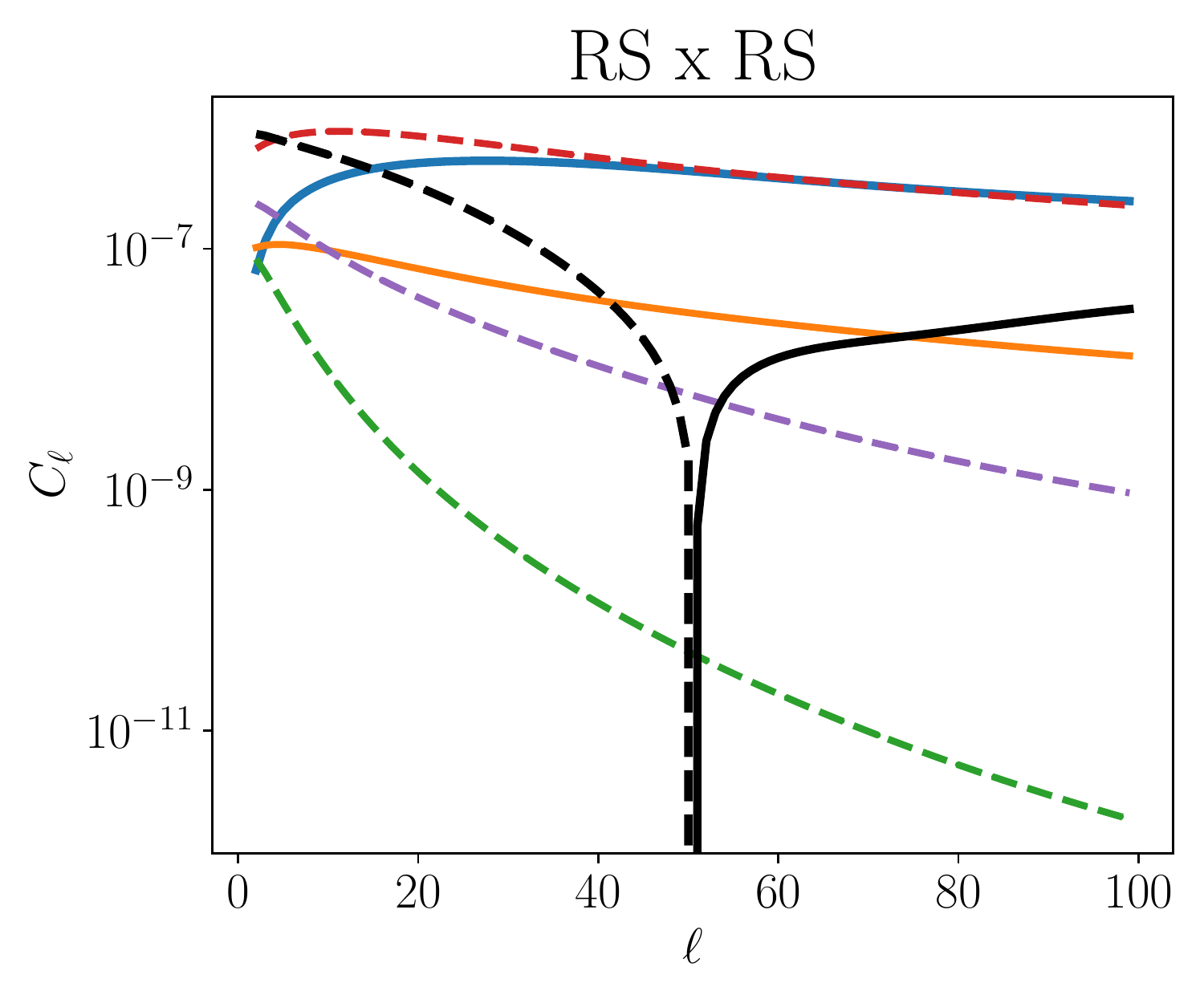}

    \includegraphics[width=0.48\textwidth]{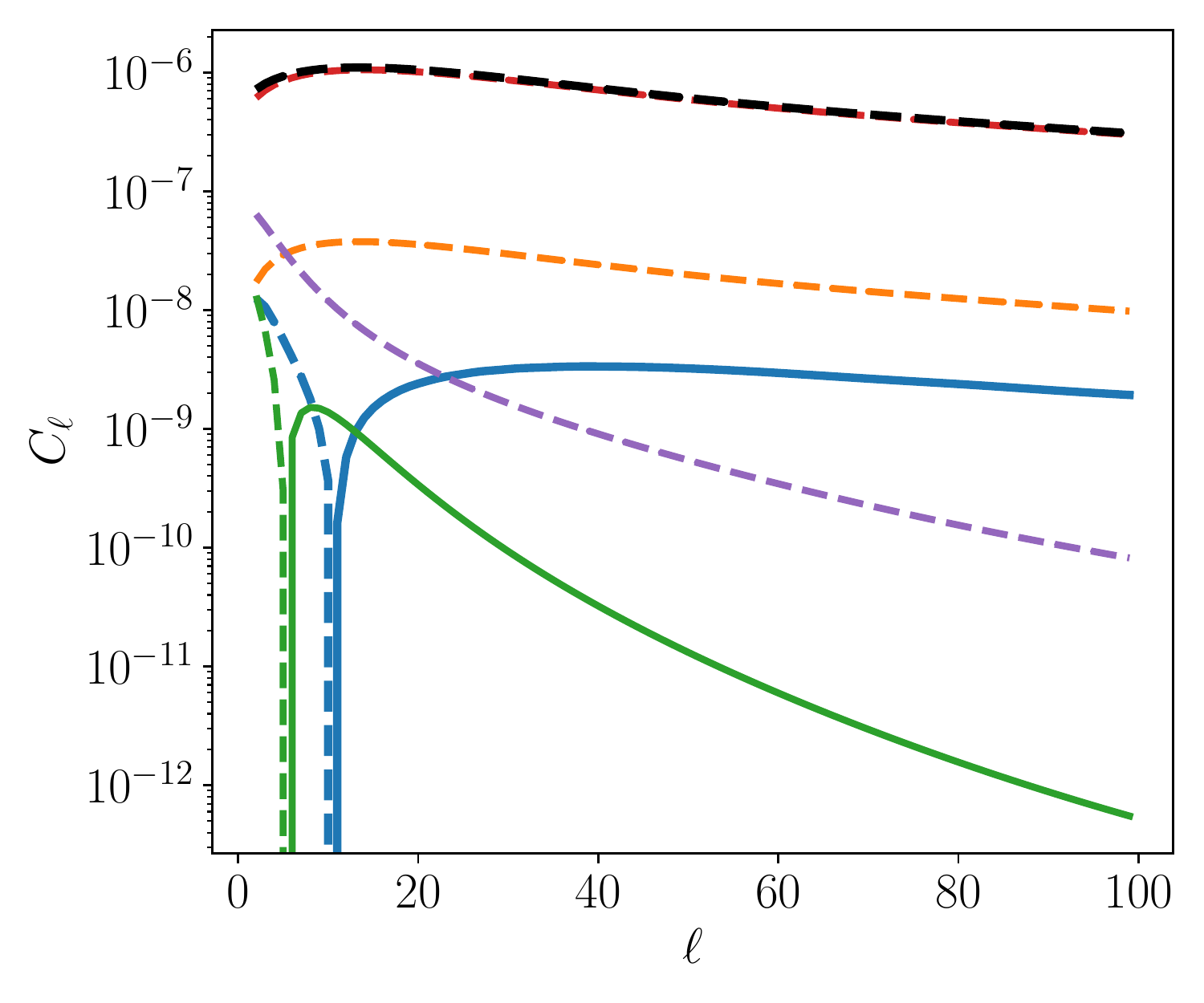}
    \includegraphics[width=0.48\textwidth]{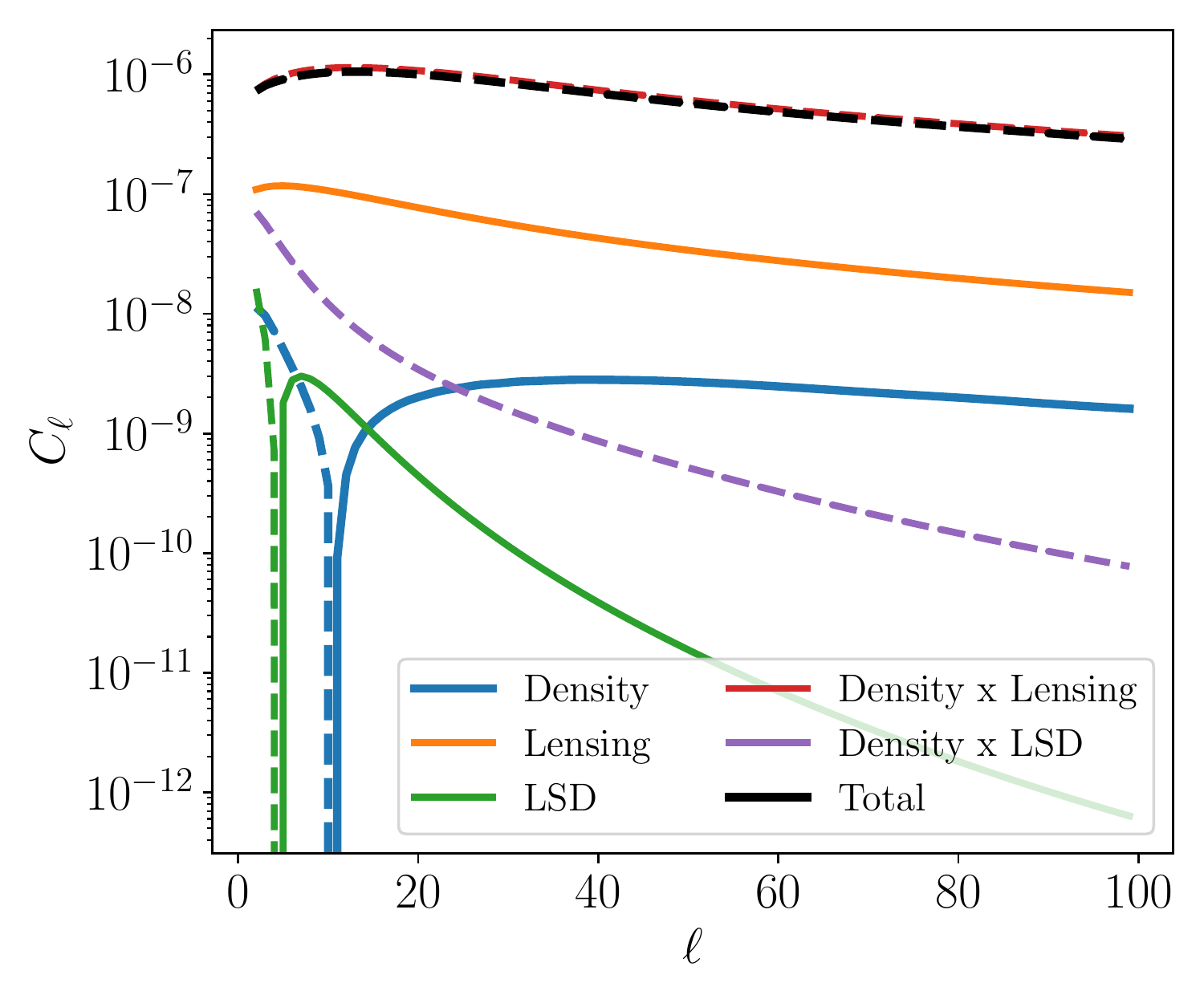}
    
    \caption{Comparison of the most important contributions (both auto- and cross-terms) when cross-correlating different redshift bins. We fix a redshift tracer, e.g. a {generic} galaxy {survey}, in the foreground at $z=0.5$, and a tracer in LDS ({\em left}) or in RS ({\em right}) in the background. The latter is set to $z=1$ for the top plots and $z=1.5$ for the bottom ones. }
    \label{fig:cross_bin}
\end{figure}
Finally, we show an example of cross-correlating two different redshift bins (see figure \ref{fig:cross_bin}). We first fix a tracer in LDS at a higher bin and cross-correlate with a lower bin in redshift space. This is what is shown on the left column of figure \ref{fig:cross_bin} together with the contributions to the cross-bin angular power spectrum in $D_L$-space, including the main auto- and cross-term correlations. The foreground survey is assumed to be a redshift space galaxy survey, as these are the most widely available. We then repeat this by setting both bins to redshift space, which is shown on the right column of figure \ref{fig:cross_bin}. The former could represent GW/SNIa signals in the background distorted by a galaxy in front, while the latter is the conventional case with a galaxy in the background and another one in front. Whilst we fix the foreground tracer at $z=0.5$, we show plots with the background one at $z=1$ (top plots) and $z=1.5$ (bottom plots). We include the three main terms and their correlations with the density term. Interestingly, \vnd{the relevant contributions to the angular power spectra on the top panels suppress} the density term on very large scales and therefore the total angular power spectrum. In all the plots we see that the density x lensing term is a key contributor to the total power, especially on very large scales. In addition the lensing-lensing correlation becomes ever more important as the redshift of the background survey increases. Although this is both true for the RS x LDS and RS x RS, the sign of these two contributions is not the same in the two cases. This is reflected in the amplitude of the cross-bin angular power which can be more than an order of magnitude different. Therefore, this reinforces the notion that when using LSS tracers, we cannot model the angular power spectrum in redshift space as a proxy to the one in luminosity distance space. One should also note that the shape and amplitude of the lines depends on the right choice of evolution and magnification biases, which we fix to zero in all cases.

\subsection{Importance of the relativistic corrections in a cosmic variance limited survey}

\begin{figure}
    \centering
    \includegraphics[width=\textwidth]{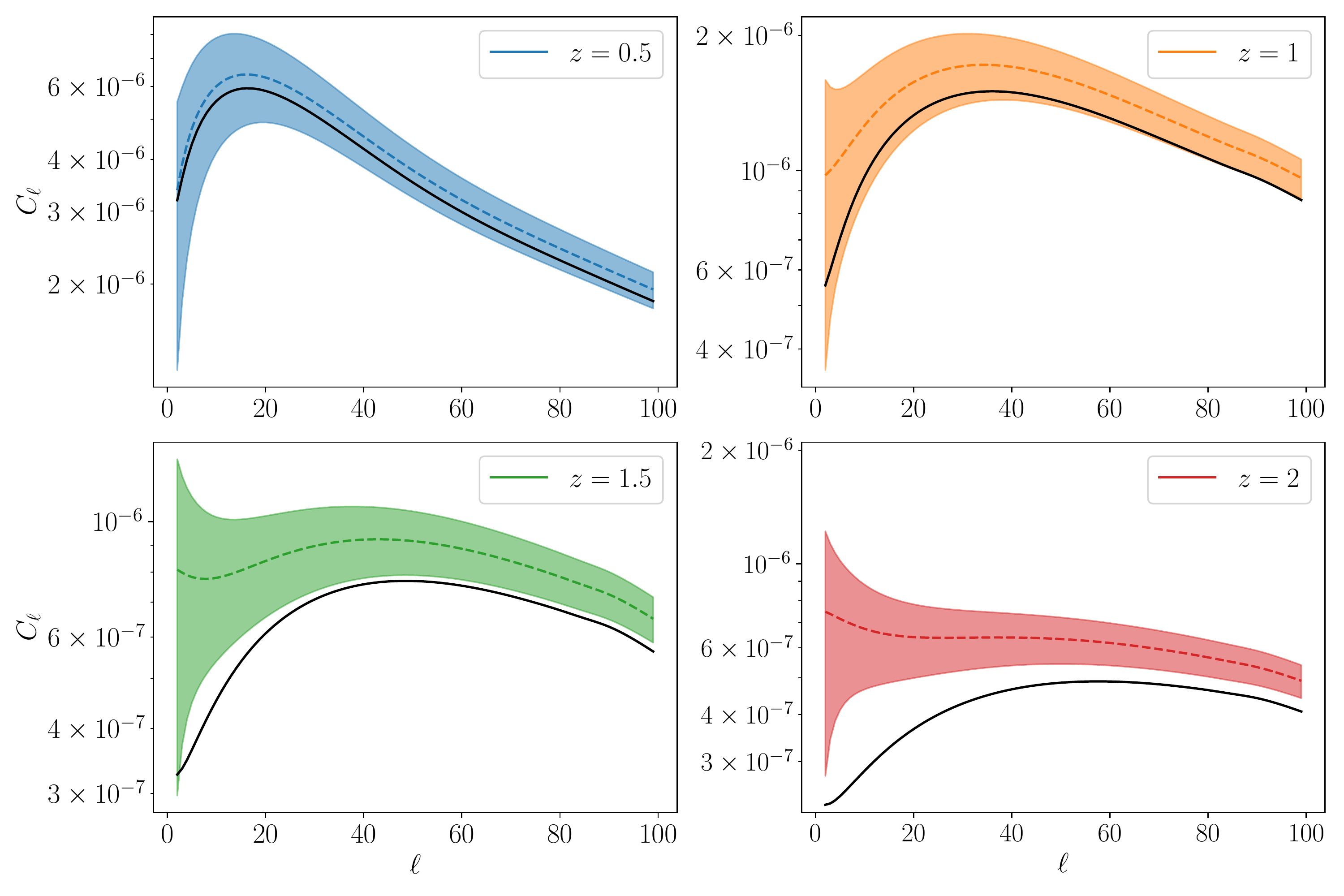}
    \caption{Error in measuring {a LDS} $C_{\ell}$ at different redshifts. In black we plot the density auto-correlation as reference.}
    \label{fig:variance}
\end{figure}

We have computed the full expression for the number density contrast in luminosity distance space including all relativistic effects. However, we have yet to show the relevance of this in the observed signal. Our goal in this subsection is not to provide any forecast on the detectability of the angular power spectrum of a source survey in luminosity distance. Neither do we intend to compare future surveys to determine which should include GR corrections, as we will focus on this in upcoming work. We therefore look at the best case scenario, i.e., {an unrealistic} cosmic variance limited survey.

The error in measuring the angular power spectrum $C_\ell$ is simply given by 
\be
\Delta C_\ell = \sqrt{\frac{2}{{(2\ell+1)}f_{\rm sky}}}\left(C_\ell+\mathcal{N}_\ell \right)\,,
\ee
where $f_{\rm sky}$ is the fraction of the observed sky and $\mathcal{N}_\ell$ is the noise. {For now, let us assume the best case scenario, even if unrealistic, and take the cosmic variance limit, where shot-noise is much smaller than the signal,} and a full sky survey as well ($f_{\rm sky}=1$). %This is the best possible scenario. 
We plot the angular power spectrum and its variance {for this idealised case} in figure \ref{fig:variance} at different redshifts. We also plot the density term only in solid black. Although we are working within an idealised scenario, one can see that at low redshifts all the correction terms would be well within the uncertainty, i.e., we may never be able to distinguish the different contributions at low z. {In this case, one can neglect the relativistic corrections, as any realistic sample will have shot-noise, which increases the uncertainty on the angular power spectrum. Furthermore, any realistic experiment will introduce sample selection uncertainties which further increase the error budget. However, as} redshift increases this is no longer true, {and the signal cannot be explained by the density term only. For redshifts $z\gtrsim 1$ the difference between the density term and the GR corrections becomes ever bigger. Therefore, as} long as the survey is not shot-noise dominated, one is required to model the signal properly. {Which terms have to be considered and the impact of neglecting them on the estimation of parameters can only be assessed for a specific experiment, including the target sample and any selection effects. It is beyond the scope of this paper as we only wanted to create a bottom line of the relevance of the GR corrections in LDS. One should not that neglecting relativistic correction at low redshifts in the cosmic variance limit would only be possible in a single tracer case, since one can go beyond cosmic variance using multiple large-scale structure tracers \cite{Fonseca:2015laa,Viljoen:2021ypp}. }

%{Note that an idealised assumpteion and in a realistic survey shot-noise is non-negligeble, neither uncertainties coming from sample selection. Here we take the best possible scenario, even if unrealistic, to create a bottom line of the relevance of the GR corrections in LDS. If, in this case they are well within the cosmic variance limit, then one could safely neglect them always.}

\section{Summary and conclusions}\label{sec:disc}

With the advent of future gravitational wave detectors \cite{Punturo_2010,Evans_2021} and optical surveys which will detect many more supernovae \cite{LSST_book}, we can start to envisage using such transient objects as tracers of large-scale structure. Observationally one can only determine the luminosity distance to these objects. Therefore, in this paper, we explored clustering in luminosity distance space. We started by computing the number density contrast in luminosity distance space in terms of the volume and luminosity distance perturbations. We also defined the magnification and evolution biases for the typical astrophysical objects which live in luminosity distance space. We then computed the luminosity distance perturbation with respect to the background redshift as well as the volume perturbation. {We note that our result agrees with \cite{Sasaki:1987ad,Pyne:2003bn,Hui:2005nm}.} We finally computed the full expression for the number counts fluctuation including all relativistic effects in LDS presented in eq. \eqref{eq:finalAs}, which is the main result of this paper. This result agrees with a previous calculation from \cite{Namikawa}{, and includes more terms than previously considered \cite{Namikawa,Zhang:2018nea,Libanore2021,Libanore2022}. We therefore provide an alternative full derivation} for the number density contrast in luminosity distance space. {We only considered transient events in luminosity distance, but our calculation is in principle valid for the stochastic gravitational background with an appropriate magnification bias.}

In deriving the density contrast we also identified how the effect of the evolution of the sources across redshift (the evolution bias) and perturbations in the number of detectable sources (the magnification bias) should be included in the calculation of the number density contrast. We derived the effect of the latter for both SNIa and GWs, considering magnitude- and signal-to-noise-limited surveys, respectively. Detailed modelling of these biases for realistic GWs and SNIa surveys, including their effects on the number counts are left for upcoming work \cite{Zazzera_2022}.

The full expression in LDS is significantly more complex than that in RS, with contributions such as lensing including additional terms. A comparison between the amplitude of each contribution is shown explicitly in \autoref{tab:table} in appendix \ref{sec:table}. Some terms behave substantially differently, such as the Doppler and the lensing ones. In particular, the Doppler term does not \vnd{diverge closer to the observer}. One surprising aspect of clustering in luminosity distance space is the sensitivity to the Doppler effect which comes via the redshift perturbation. 
The area distance perturbation is a purely geometric effect giving rise to the normal lensing contribution in the Jacobi map~\cite{Bacon2014}. When moving to luminosity distance the reciprocity relation brings in the redshift, and therefore the Doppler contribution to the redshift perturbation. 

We then explored the angular power spectrum in LDS. To do so, we changed the publicly available cosmological solver \texttt{CAMB}, implementing the luminosity distance number counts transfer functions. The code not only computes the angular power for luminosity distance sources, but also their correlation with other tracers of dark matter such as galaxy surveys. The code will be made available in the future. With this tool, we found a significant percentage difference with the one in redshift space at large scales. At higher redshifts, this difference increases and can reach \vnd{$20\%$} at $z=2.5$. This is extremely important as future gravitational wave detectors have large effective beams, meaning that only low $\ell$ multipoles would be accessible; these are the same multipoles where the number density contrast in redshift space is not a good approximation of the one in luminosity distance space. This should have a significant impact on the detectability and best-fit values of the astrophysical biases of gravitational wave mergers. These in turn can be used to characterise the distribution and evolution of black holes with cosmic time.

A notable difference between the two spaces is the importance of lensing, especially in cross-correlations.
Cross-correlating different redshift bins shows different behaviours of the lensing terms, namely the density x lensing contribution. In the case examined, whilst the angular power spectrum in LDS is fully positive at higher redshifts, in redshift space it crosses zero at large scales. 

Therefore, mismodelling the angular power spectra when cross-correlating a background GWs survey with a foreground lensing object, e.g. a galaxy survey, will lead to a misunderstanding of the gravitational potentials in the line-of-sight. In fact, if we attain a cosmic variance limited sample in the future, not properly including all GR effects will lead to a total mismatch between the observed and the density-only contribution (as we have shown in figure \ref{fig:variance}).

Several works in the literature have studied how transient events, such as GW mergers and SNIa, can be used as tracers of the dark matter distribution in the universe. Our contribution to this debate was centered in providing a full calculation of the number density contrast, including all GR effects, in the space where distances are observed. In contrast with galaxy surveys, here we estimate luminosity distances rather than redshifts. We have shown that this difference is substantial and should be included in the modelling and forecasts.

\acknowledgments
We are pleased to thank Charles Dalang, Roy Maartens and Stefano Camera for useful discussions. \vnd{We also thank Anna Balaudo, Mattia Pantiri and Alessandra Silvestri for checking our derivation and pointing out one minus sign that wasn't carried on properly.} JF thanks the support of Funda\c{c}\~{a}o para a Ci\^{e}ncia e a Tecnologia (FCT) through the
research grants UIDB/04434/2020 and UIDP/04434/2020 and through
the Investigador FCT Contract No. 2020.02633.CEECIND/ CP1631/CT0002. JF also thanks the hospitality of Astronomy Unit of QMUL and the University of the Western Cape where part of this work was developed. We thank the support of FCT and the Portuguese Association of Researcher and Students in the UK (PARSUK) under the Portugal - United Kingdom exchange program Bilateral Research Fund (BRF). T.B. is supported by ERC Starting Grant \textit{SHADE} (grant no.~StG 949572) and a Royal Society University Research Fellowship (grant no.~URF$\backslash$R1$\backslash$180009). CC is supported by the UK Science \& Technology Facilities Council Consolidated Grant ST/P000592/1. S.Z. acknowledges support by the Perren Fund University of London.

\newpage
\appendix

\section{The luminosity distance density contrast in harmonic space}\label{sec:LDharm}

In section \ref{sec:expression} we presented the number density contrast in luminosity distance space (see eq. \ref{eq:num_final}). In practice one looks at the angular distribution of sources and the statistical structure of its distribution to relate it with cosmological parameters and the dark matter density contrast. In this appendix we will revise the multipole expansion in harmonic space and compute the angular power spectrum $C_{\ell}$. We will also show how we derive the transfer functions in harmonic space and summarise our implementation in \texttt{CAMB}.

\subsection{Decomposition into spherical harmonics}\label{sec:spherical}

By definition the number density contrast is a perturbed quantity with spatial average $\langle \Delta\rangle=0$. Here we work in shells in the celestial sphere and, thus, do a spherical harmonic decomposition of the field $\De(\bn,D_L)$ into
\be
\De(\bn,D_L)= \sum^{\infty}_{\ell = 0} \sum^{\ell}_{m = -\ell} \alm(D_L) \ylm{\bn}\,.
\ee
The only dependence on the distance is encoded in the $\alm$, which are the projection of the field onto the spherical harmonic basis.
It follows that the $\alm$ can then be written in terms of the field as (see appendix \ref{sec:propSH})
\be \label{eq:alm_delta}
\alm(D_L)=\int \d\Omega_{\bn}\ \De(\bn,D_L) \ylmc{\bn} \, .
\ee
The $\alm$ will keep the same statistical properties of $\De$, i.e., $\langle \alm \rangle = 0$. It is in effect a random variable with null average where the 2-point function describes its statistical distribution. It is this 2-point function of the $\alm$ which we call the angular power spectra $C_\ell$, and is defined by
\be
\langle \alm a^*_{\ell'm'} \rangle \equiv C_\ell\ \delta_{\ell\ell'}\delta_{mm'}\,.
\ee
So far we have assumed an infinitesimal shell in $D_L$, which in practice will never happen. In reality, data is binned into intervals of distance and the ``observed'' $\alm^i$ in the i\emph{th}-bin is then a weighted quantity, i.e.,
\be \label{eq:almbin}
\alm^i(\bar D_L)=\int {\rm d}D_L\ w_i(D_L)\ \alm(D_L) \,,
\ee
where $w_i(D_L)$ is the normalized distribution function and $\bar D_L$ is the central distance of the bin.
The weighting is usually given both by the distribution of sources $\bar n$ along the line-of-sight and by the window function $W_i(D_L)$ we have chosen, i.e. how we bin the tracers considered. The window function can be e.g. a top-hat, a gaussian or an error function and will depend on the type of survey. In all our analysis we opted for broad gaussian windows. In general one can take the weighting function to be 
\be
w(D_L)=\frac1{\int\d D_L'\ W_i(D_L') \bar n(D_L')} W_i(D_L) \bar n(D_L)\,.
\ee

\subsection{The angular power spectra}

Let us consider for now the density term $\delta_M$. Note that the bias is in this case an overall amplitude so we can neglect it for now. On what follows the comoving distance is implicitly defined as a function of the luminosity distance, i.e., $r\equiv r(D_L)$. We can write the density $\alm$ as
\bea
\alm^\delta(D_L) &=& \int\ \d\Omega_{\bn}\ \delta_M(\bn,r) \ylmc{n}\nn\\
&=&\frac1{(2\pi)^3} \int\int \d\Omega_{\bn}\ \dt k\ \delta_{\vk}(D_L)\  \ylmc{n}\ e^{i kr \hat k.\bn}\nn\\
&=&\frac1{(2\pi)^3} \int\int \d\Omega_{\bn}\ \dt k\ \delta_{\vk}(D_L)\  \ylmc{n}\ \sum_{\ell'=0}^{\infty}(2\ell'+1)i^{\ell'} {\cal P}_{\ell'}(\hat k.\bn) j_{\ell'}(kr)\nn\\
&=&\frac{i^{\ell}}{2\pi^2} \int \dt k\ \delta_{\vk}(D_L)\  \ylmc{k}\ j_{\ell}(kr)
\,.
\eea
In the second line we Fourier transform $\delta_M$, while in the third and fourth lines we used properties of Legendre polynomials and spherical harmonics (see appendix \ref{sec:propSH}). Then, using Eq. \ref{eq:almbin}, the two point function between two bins is
\bea
\left\langle \alm^{\delta}(D_{L,i}) a^{\delta*}_{\ell'm'}(D_{L,j}) \right\rangle&=&\Bigg\langle \left[\int {\rm d}D_L\ w^i(D_L)\frac{i^{\ell}}{2\pi^2} \int \dt k\ \delta_{\vk}(D_L)\  \ylmc{k}\ j_{\ell}(kr)\right]\times\nn\\
&&\left[\int {\rm d}D_L'\ w^j(D_L')\frac{(-i)^{\ell'}}{2\pi^2} \int \dt k'\ \delta^*_{\vk'}(D_L')\  Y_{l'm'}(\hat k')\ j_{\ell'}(k'r'))\right]\Bigg\rangle \nn\\
&=&\frac{(-1)^{\ell'}i^{\ell+\ell'}}{4\pi^4}  \int {\rm d}D_L\ w^i(D_L)\int \dt k\ \ylmc{k}\ j_{\ell}(k r)\times\nn\\
&&\int {\rm d}D_L'\ w^j(D_L') \int \dt k'\ Y_{l'm'}(\hat k')\ j_{\ell'}(k'r') \times \nn\\
&&\left\langle \delta_{\vk}(D_L) \delta^{*}_{\vk'}(D_L') \right\rangle \label{eq:cl_density_inter}
\eea
%\nn\\
%&&
Here it is useful to introduce transfer function that split the redshift evolution from the primordial inflationary perturbation. We refer to these as
\be
{\cal T}_{\cal Q}(D_L,k)\equiv\frac{{\cal Q}(D_L,k)}{{\cal R}(D_{L,initial},k)}\,,
\ee
where
\be
\langle {\cal R}(k) {\cal R}^*(k')\rangle=(2\pi)^3\delta^3(\vec k-\vec k') P_{\cal R}(k)\,,
\ee 
and $P_{\cal R}(k)$ is the power spectrum primordial perturbations and $\cal R$ is the curvature perturbation. This way
\be
\langle {\cal Q}(D_L,k) {\cal Q}^*(D_L',k')\rangle=(2\pi)^3\delta^3(\vec k-\vec k') {\cal T}_{\cal Q}(D_L,k) {\cal T}_{\cal Q}(D_L',k) P_{\cal R}(k)\,.
\ee
The primordial power spectrum is given by
\be
{\cal P}_{\cal R}(k)  \equiv \frac{k^3}{2\pi^2}P_{\cal R}(k)=A_{\cal S}\left(\frac{k}{k_p}\right)^{n_{\cal S}-1}\,,
\ee
where $k_p$ is a pivot scale and $A_{\cal S}$ is the primordial amplitude of the power spectrum and $n_{\cal S}$ its spectral index. With this in mind we can then write
\be
\left\langle \delta_{\vk}(D_L) \delta^*_{\vk'}(D_L') \right\rangle = (2\pi)^3\delta^3(\vec k-\vec k') {\cal T}_{\cal \delta}(D_L,k) {\cal T}_{\cal \delta}(D_L',k) P_{\cal R}(k)\,.
\ee
We can then simplify eq. \ref{eq:cl_density_inter} as 
\bea
\left\langle \alm^{\delta}(\bar D_{L,i}) a^{\delta*}_{\ell'm'}(\bar D_{L,j}) \right\rangle&=&\frac{2(-1)^{\ell'}i^{\ell+\ell'}}{\pi} \left(\int {\rm d}\Omega_{\hat k}\  \ylmc{k}Y_{l'm'}(\hat k) \right)  \int k^2 {\rm d}k\ P_{\cal R}(k) \nn\\
&& \left[\int {\rm d}D_L\ w^i(D_L) {\cal T}_{\cal \delta}(D_L,k)\ j_{\ell}(kr)\right] \left[\int {\rm d}D_L'\ wj(D_L'){\cal T}_{\cal \delta}(D_L',k)\ j_{\ell'}(kr')\right] \nn\\
&=&4\pi \int {\rm d}\ln k\ \Delta^{\delta}_{\ell}(k,\bar D_{L,i})\Delta^{\delta}_{\ell}(k,\bar D_{L,j}) {\cal P}_{\cal R}(k)~ \delta_{\ell\ell'}\delta_{mm'}\,,
\eea
where we implicitly define the effective transfer function $\Delta^{\delta}$ as
\be
\Delta^{\delta}_{\ell}(k,\bar D_{L,i})=\int_0^{\infty}  {\rm d}D_L\ w^i(D_L) b(D_L) {\cal T}_{\cal \delta}(D_L,k)\ j_{\ell}(kr)\,,
\ee
where we reintroduced the bias $b$. Note that this derivation holds for the potentials as well, but not for their gradient or the velocity terms. These include a derivative term with respect to $k$ (in Fourier space), which affects the integrand of the transfer function (see e.g. $\Delta_{\ell}^{\nabla\Phi}$ and $\Delta_{\ell}^{LSD}$ in appendix \ref{sec:transfer}).
As \texttt{CAMB} is written in conformal time $\eta$ it is more convenient to present $\Delta^{\delta}$ as
\be
\Delta^{\delta}_{\ell}(k,D_{L,i})=-\int^{\eta_0}_{0} {\rm d}\eta\ \frac{\d D_L}{\d\eta} w^i(\eta) b(\eta){\cal T}_{\cal \delta}(\eta,k) j_{\ell}(k(\eta_0-\eta))\,,
\ee
where in units of $c=1$, $r=\eta_0-\eta$, and
\be\label{eq:dDL_deta}
\frac{\d D_L}{\d\eta} = -\frac{1+(\eta_0-\eta)\cal H}{a}\,.
\ee
The angular power spectrum is then given by
\be
C^{ij}_\ell= 4\pi \int {\rm d}\ln k\ \De^i_{\ell}(k,D_{L,i})\De^j_{\ell}(k,D_{L,j})\ {\cal P}_{\cal R}(k)\,,
\ee
with
\be
\De_{\ell}=\De^{\delta}_{\ell}+\De^{LSD}_{\ell}+\De^{L}_{\ell}+\De^{D}_{\ell}+\De^{\Phi}_{\ell}+\De^{\Psi}_{\ell}+\De^{\Phi'}_{\ell}+\De^{\nabla\Phi}_{\ell}+\De^{TD}_{\ell}+\De^{ISW}_{\ell}+\De^{\delta_v}_{\ell}\,.
\ee

\newpage

\section{Transfer functions implemented in \texttt{CAMB}}\label{sec:transfer}
The transfer functions implemented are given by
\bea
\Delta_\ell^{\delta} &=& \int_0^{\eta_0} \d\eta\ 
\left|\frac{\d D_L}{\d\eta}\right|w(\eta)b(\eta){\cal T}_{\delta}(\eta,k)j_\ell(k(\eta_0-\eta))\,,\\
\Delta_\ell^{LSD} &=& - \int_0^{\eta_0} \d\eta\ \frac{\d^2}{\d^2\eta} \left[ \left|\frac{\d D_L}{\d\eta}\right| w(\eta) \frac{A^{LSD}(\eta)}{k} {\cal T}_{v}(\eta,k)\right] j_\ell(k(\eta_0-\eta)) \,,\\
\Delta_\ell^{L} &=& \vnd{-}\ell(\ell+1) \int^{\eta_0}_0 \d\eta'\ \left({\cal T}_{\Phi}(\eta',k)+{\cal T}_{\Psi}(\eta',k)\right) j_{\ell}(k(\eta_0-\eta')) \times \nn\\
&&\int^{\eta'}_0 \d \eta\  \left|\frac{\d D_L}{\d\eta}\right| \frac{w_i(\eta)}{\vnd{(\eta_0-\eta)}} A_L(\eta,\eta')\,,\\
\Delta_\ell^{D} &=& \int_0^{\eta_0} \d\eta\ \frac{\d}{\d\eta} \left[ \left|\frac{\d D_L}{\d\eta}\right| w(\eta) \frac{A^{D}(\eta)}{k} {\cal T}_{v}(\eta,k)\right] j_\ell(k(\eta_0-\eta))\,,\\
\Delta_\ell^{\Phi} &=& \int_0^{\eta_0} \d\eta\ 
\left|\frac{\d D_L}{\d\eta}\right|w(\eta)A^{\Phi}(\eta){\cal T}_{\Phi}(\eta,k)j_\ell(k(\eta_0-\eta))\,,\\
\Delta_\ell^{\Psi} &=& \int_0^{\eta_0} \d\eta\ 
\left|\frac{\d D_L}{\d\eta}\right|w(\eta)A^{\Psi}(\eta){\cal T}_{\Psi}(\eta,k)j_\ell(k(\eta_0-\eta))\,,\\
\Delta_\ell^{\Phi'} &=& \int_0^{\eta_0} \d\eta\ 
\left|\frac{\d D_L}{\d\eta}\right|w(\eta)A^{\Phi'}(\eta){\cal T}_{\Phi'}(\eta,k)j_\ell(k(\eta_0-\eta))\,,\\
\Delta_\ell^{\nabla\Phi} &=& \int_0^{\eta_0} \d\eta\ \frac{\d}{\d\eta} \left[ \left|\frac{\d D_L}{\d\eta}\right| w(\eta) A^{\nabla \Phi}(\eta) {\cal T}_{\Phi}(\eta,k)\right] j_\ell(k(\eta_0-\eta))\,,\\
\Delta_\ell^{TD} &=& \int^{\eta_0}_0 \d\eta'\ \left({\cal T}_{\Phi}(\eta',k)+{\cal T}_{\Psi}(\eta',k)\right) j_{\ell}(k(\eta_0-\eta'))\times \nn\\
&& \int^{\eta'}_0 \d \eta\  \left|\frac{\d D_L}{\d\eta}\right| \frac{w_i(\eta)}{\vnd{(\eta_0-\eta)}} A_{TD}(\eta,\eta')\,,\\
\Delta_\ell^{ISW} &=& \int^{\eta_0}_0 \d \eta'\ \left({\cal T}_{\Phi'}(\eta',k)+{\cal T}_{\Psi'}(\eta',k)\right)  j_{\ell}(k(\eta_0-\eta'))\times \nn\\
&&\int^{\eta'}_0 \d \eta\ \left|\frac{\d D_L}{\d\eta}\right| w_i(\eta) A_{\rm ISW}(\eta)\,,\\
\Delta_\ell^{\delta_v} &=& \int_0^{\eta_0} \d\eta\ 
\left|\frac{\d D_L}{\d\eta}\right|w(\eta)\left[b_e-3\right]\frac{\mathcal{H}}{k}{\cal T}_{v}(\eta,k)j_\ell(k(\eta_0-\eta))\,.
\eea

\newpage

\section{Properties of spherical harmonics}\label{sec:propSH}

The spherical harmonics $\ylm{n}$ obey the differential equation
\be \label{eq:defsh}
\Delta_{\bn} \ylm{\bn}=-\ell(\ell+1)\ylm{\bn}\,,
\ee
where
\be
\Delta_{\bn}=\frac1{\sin\theta}\frac{\partial}{\partial \theta}\left(\sin\theta\frac{\partial}{\partial \theta}\right) + \frac1{\sin^2\theta}\frac{\partial^2}{\partial\varphi^2}\,.
\ee
The spherical harmonics are a natural set of base functions for mode decomposition on the sphere as they are orthonormal, i.e.
\be \label{eq:sh:norm}
\int \d\Omega_\bn\ \ylm{\bn}Y^*_{\ell'm'}(\bn ) = \delta_{\ell\ell'}\delta_{mm'}\,.
\ee
They are also related to the Legendre polynomials by
\be \label{eq:sh:leg}
{\cal P}_\ell(\bn\cdot\bn') = \frac{4\pi}{2\ell+1}\sum^{m=\ell}_{m=-\ell} \ylm{\bn} \ylmc{\bn'}\,.
\ee
It follows from eqs. (\ref{eq:sh:norm}) and (\ref{eq:sh:leg}) the useful expression
\be \label{prop:Ylm_Leg}
\int \don\ {\cal P}_{\ell'}(\bn\cdot\bn')\  \ylmc{\bn}\ = \frac{4\pi}{2\ell+1}\ylmc{\bn'} \delta_{\ell\ell'}\,, 
\ee
which we will use regularly, and which is valid for the complex conjugate. Another useful property is 
\be \label{prop:FT_LegBessel}
e^{i r \bn\cdot\bn'}=\sum_{\ell=0}^{\infty}(2\ell+1)\ i^\ell\ {\cal P}_\ell(\bn\cdot\bn')\ j_\ell(r)\,,
\ee
where $j_\ell$ are the spherical Bessel functions. Then
\bea
\int \don\ e^{i r \bn\cdot\bn'} Y^*_{\ell m}(\bn)&=& \int \don\ \sum_{\ell'=0}^{\infty}(2\ell'+1)\ i^{\ell'}\ {\cal P}_{\ell'}(\bn\cdot\bn')\ j_{\ell'}(r) Y^*_{\ell m}(\bn)\nn\\
&=& \sum_{\ell'=0}^{\infty}(2\ell'+1)\ i^{\ell'}\ j_{\ell'}(r) \frac{4\pi}{2\ell+1}\ylmc{\bn'} \delta_{\ell\ell'} \nn\\
&=& 4\pi\ i^{\ell}\ j_{\ell}(r) \ylmc{\bn'} \label{eq:int_ylms_ft}
\eea
Finally, a further property that will be useful follows from eqs. (\ref{eq:defsh}) and (\ref{eq:sh:leg}), i.e. 
\be \label{eq:lapleg}
\Delta_{\bn} {\cal P}_{\ell}(\bn\cdot\bn')=-\ell(\ell+1){\cal P}_{\ell}(\bn\cdot\bn')\,.
\ee

\newpage
\begin{landscape}
\section{Comparison of the amplitudes}\label{sec:table}

\begin{table}[h] 
\centering
\begin{tabular}{|c|c|c|}
\toprule
\textbf{Term} & \textbf{LDS} & \textbf{RS} \\
\midrule
 $A_{LSD,RSD}$ & $-\frac{2\bar r}{1+\bar{r}\mathcal{H}}$ & $-\frac1{\mathcal{H}}$ \\
 \midrule
$A_D$ & $\vnd{-1+10s-\frac{2\bar{r}\mathcal{H}}{1+\bar{r}\mathcal{H}}\left(\frac2{\bar{r}\mathcal{H}}-b_e\right) }-2\left(\frac{\bar{r}\mathcal{H}}{1+\bar{r}\mathcal{H}}\right)^2\left(\frac{\mathcal{H}'}{\mathcal{H}^2}-\frac1{\bar{r}\mathcal{H}}\right)$ & $\frac{5s-2}{\bar{r}\mathcal{H}}-5s+b_e-\frac{\mathcal{H}'}{\mathcal{H}^2}$ \\
\midrule
$A_{\Psi}$ & $\vnd{3-10s}+\frac{\bar{r}\mathcal{H}}{1+\bar{r}\mathcal{H}}\left(\frac3{\bar{r}\mathcal{H}}-2-2b_e\right)+2\left(\frac{\bar{r}\mathcal{H}}{1+\bar{r}\mathcal{H}}\right)^2\left(\frac{\mathcal{H}'}{\mathcal{H}^2}-\frac1{\bar{r}\mathcal{H}}\right)$ & $1+\frac{5s-2}{\bar{r}\mathcal{H}}-5s+b_e-\frac{\mathcal{H}'}{\mathcal{H}^2}$ \\
\midrule
$A_{\Phi}$ & $\vnd{-1-5s}+\frac{\bar{r}\mathcal{H}}{1+\bar{r}\mathcal{H}}\left(\frac1{\bar{r}\mathcal{H}}-1-b_e\right) +\left(\frac{\bar{r}\mathcal{H}}{1+\bar{r}\mathcal{H}}\right)^2\left(\frac{\mathcal{H}'}{\mathcal{H}^2}-\frac1{\bar{r}\mathcal{H}}\right)$ & $5s-2$ \\
\midrule
$A_{\Phi'}$ & $\frac{\bar{r}}{1+\bar{r}\mathcal{H}}$ &$\frac1{\mathcal{H}}$\\
\midrule
$A_{\nabla\Phi}$ & $\frac{\bar{r}}{1+\bar{r}\mathcal{H}}$ & $0$ \\
\midrule
$A_{TD}$ & $\vnd{1+5s-\frac{\bar{r}\mathcal{H}}{1+\bar{r}\mathcal{H}}\left(\frac1{\bar{r}\mathcal{H}}-1-b_e\right)}-\left(\frac{\bar{r}\mathcal{H}}{1+\bar{r}\mathcal{H}}\right)^2\left(\frac{\mathcal{H}'}{\mathcal{H}^2}-\frac1{\bar{r}\mathcal{H}}\right)$ & $5s-2$ \\
\midrule
$A_L$ & $\frac1{2}\left[\left(\frac{\bar{r}-r}{r}\right)\left(\vnd{-1-5s}+\frac{\bar{r}\mathcal{H}}{1+\bar{r}\mathcal{H}}\left(\frac2{\bar{r}\mathcal{H}}-1-b_e\right)+\left(\frac{\bar{r}\mathcal{H}}{1+\bar{r}\mathcal{H}}\right)^2\left(\frac{\mathcal{H}'}{\mathcal{H}^2}-\frac1{\bar{r}\mathcal{H}}\right)\right)+\frac1{1+\bar{r}\mathcal{H}}\right]$ & $\frac1{2}(5s-2)\frac{\bar{r}-r}{r}$ \\
\midrule
$A_{ISW}$ & $\vnd{2-10s+\frac{\bar{r}\mathcal{H}}{1+\bar{r}\mathcal{H}}\left(\frac4{\bar{r}\mathcal{H}}-2-2b_e\right)+2\left(\frac{\bar{r}\mathcal{H}}{1+\bar{r}\mathcal{H}}\right)^2\left(\frac{\mathcal{H}'}{\mathcal{H}^2}-\frac1{\bar{r}\mathcal{H}}\right)}$ & $\frac{2-5s}{\bar{r}\mathcal{H}}+5s-b_e+\frac{\mathcal{H}'}{\mathcal{H}^2}$ \\
\bottomrule
\end{tabular}
\caption{Comparison of the amplitudes of the corrections to the number counts fluctuation between luminosity distance space and redshift space.} \label{tab:table}
\end{table}
\end{landscape}

\end{document}